\documentclass[pre,
preprint,
showpacs,preprintnumbers,amsmath,amssymb]{revtex4-1}
\usepackage{graphicx}
\usepackage{epstopdf}
\usepackage{dcolumn}
\usepackage{bm}
\usepackage{latexsym} 
\usepackage{amsmath} 
\usepackage{natbib} 
\usepackage{color}
\usepackage[usenames,dvipsnames]{xcolor}

\begin{document}

\title{Finite-time effects on a first-order irreversible phase transition}

 \author{E. S. Loscar} 
 \affiliation{%
   Instituto de F\'{i}sica de L\'{i}quidos y Sistemas Biol\'ogicos
   (IFLYSIB), UNLP, CCT La Plata-CONICET, Calle 59 no.~789, B1900BTE La
   Plata, Argentina} 
 
 \affiliation{Departamento de F\'{i}sica, Universidad Nacional de La
   Plata, c.c. 67, 1900 La Plata, Argentina}
 \email{yasser.loscar@gmail.com}
 
 \begin{abstract}
The first-order irreversible phase transition (FOIPT) of the ZGB model [Ziff, Gulari, Barshad, Phys. Rev. Lett. \textbf{56} (1986) 2553] for the catalytic oxidation of carbon monoxide is studied numerically in the presence 
of a slowly-varying, time-dependent, and spatially-uniform carbon monoxide pressure, 
by means of standard controlled-pressure simulations. 
This method allows us to observe finite-time effects close to the FOIPT, as well as evidence that a dynamic phase transition occurs. The location of this transition is measured very precisely and compared with previous results in the literature.
\end{abstract}

\maketitle

\section{\label{sec:Intro}Introduction}
Statistical systems exhibit rich off-equilibrium behaviors when one of the intensive model parameters, such as the temperature or the magnetic field in spin systems, varies smoothly across a phase-transition point with a controlled time scale $t_s$.
If $p_c$ is the critical value of the intensive parameter $p$, one considers the dynamic linear protocol (DLP) given by $p/p_c=1-t/t_s$, starting from $t=t_i<0$ to $t=t_f>0$.

These off-equilibrium phenomena have received a lot of attention in continuous phase transitions where, as in the case of the equilibrium static and dynamic critical behavior, the off-equilibrium behaviors show universal features. In the limit of very large $t_s$ and close to the transition point, it is possible to obtain general scaling relations in terms of the static and dynamic standard critical exponents \cite{kibble1976,kibble2007,zurek1985,Gong2010,Chandran2012,Zhong2011}.
A prototypical example is the Kibble-Zurek mechanism, first proposed by Kibble in cosmology \cite{kibble1976}, for the formation of topological defects when the temperature is slowly changed across a symmetry-breaking transition from the disordered phase to the ordered phase. Later, Zurek suggested that ideas about how to calculate the length correlation present in the
transition could be tested by applying them to transitions in condensed-matter physics \cite{zurek1985}.
These phenomena have been studied in different branches of condensed-matter physics, such as in liquid crystals \cite{liquidcristal54,liquidcristal55}, superconductors \cite{superconductores58}, 
superfluids \cite{superfulidos56,superfluido57}, and also at quantum phase transitions  \cite{quantum1,quantum2,quantum3,quantum4}. 
More recently, the DLP that varies the magnetic field at the critical temperature
in continuous magnetic transitions, such as those of a three-dimensional topological magnet \cite{castelnovo}
and the three-dimensional $O(N)$ vector models \cite{tanos1}, has been investigated.
It has also been used to relate finite-time scaling to the Kibble-Zurek mechanism and to demonstrate it numerically for two- and three-dimensional Ising models \cite{Zhong2014}. 

The DLP has also been applied to first-order transitions.
In the three-dimensional $O(N)$ vector models below the critical temperature, the magnetization displays an off-equilibrium scaling behavior close to the transition line at null field \cite{tanos1}. The exponents involved are different from those corresponding to the critical temperatures and depend on the shape of the lattice and the boundary conditions. 
The off-equilibrium behavior also arises in the thermal first-order transition of the two-dimensional Potts model with $q=10,20$ \cite{tanos2}. 
In this work, numerical results show evidence of a dynamic transition
where the scaling functions exhibit a spinodal-like singularity. This singular behavior resembles that at the mean-field spinodal point \cite{Gunton1978,Klein1982,Binder}. 
However, the location of this dynamic singularity converges to the transition point when the characteristic time $t_s\rightarrow \infty$. More recently, the Kibble-Zurek protocol has been applied to the first-order transition of the quantum-Ising chain \cite{tanos3,tanos4}. 

Apart from the Hamiltonian systems described above, there are far-from-equilibrium 
systems that exhibit irreversible phase transitions (IPT's) \cite{DickmanBook}.
Archetypal examples are catalytic surface reaction systems \cite{ReviewNos} and spreading processes \cite{Hinrichsen2000}, among others. 
The associated models are intrinsically non-equilibrium systems whose properties are not derived from a Hamiltonian but are fully determined by dynamic rules. 
In contrast to their equilibrium counterparts, IPT's in reactive systems occur between an active regime with a sustained outcome of the reaction products from the catalytic surface and an absorbing state in which the catalyst becomes fully covered by one or more types of reactants. Because the systems cannot escape from this absorbing state, the transitions are irreversible. In this context, the catalytic oxidation of $CO$ is one of the most studied reactions due to its practical importance and theoretical interest \cite{ReviewNos}. 
The simplest approach for this catalytic reaction is the ZGB lattice-gas model \cite{zgb}. 
The ZGB is a minimal model with only one external control parameter, which is the normalized pressure of $CO$, and it exhibits an active window between 
a second-order IPT and a first-order IPT. Despite the simplicity of the model, these transitions act in many ways like equilibrium phase transitions, exhibiting critical behavior, metastability, nucleation, hysteresis, etc. The discontinuous IPT in the ZGB model is also in qualitative agreement with the sharp reduction in the rate of $CO_2$ production observed when the $CO$ pressure is increased in the catalytic oxidation of $CO$ on Pt single-crystal surfaces \cite{Ehsasi1989,Berdau1999}. 
Furthermore, experimental evidence of hysteretic effects in this reaction has been reported \cite{Berdau1999}, supporting the first-order character of the transition. 
In agreement, mean field approaches to the ZGB model predict the first-order IPT and also the existence of metastable states and a spinodal point 
\cite{Dickman1986,evans1991,Zhdanov1994,DickmanBook}. 

Simulations of the ZGB model have been performed mainly using two different ensembles. 
These have been done extensively using the {\it standard ensemble}, in 
which the pressure of $CO$ is the control parameter \cite{zgb,ziff2010}.
The determination of the first-order transition is affected by metastability, 
which can be avoided by means of convenient initial conditions \cite{ziff1992}. 
Other simulations, such as the short-time dynamics for the determination of the spinodal \cite{albanoSTD,daSilva2016}, the periodic variation of 
the CO pressure with CO desorption to find a nonequilibrium dynamic phase transition
\cite{Rikvold2005}, and rare-event techniques to compute transition times for nucleation \cite{ziff2010}, have also been developed.
On the other hand, in the simulation using the so-called {\it constant-coverage} ensemble, the $CO$ -coverage is kept constant on average and is the control parameter \cite{ziff1992,albanoSTD,monetti2001,ccpnos,ccpnos2}.
This ensemble has been applied numerically to thoroughly study 
the first-order transition, allowing exploration of the coexistence region.
The constant-coverage ensemble permits high-precision determination of the transition point, finding evidence of the evaporation/condensation transition of droplets, determining spinodal points, studying hysteretic effects related to the curvature of the interface, among others.
The great effort dedicated to the study of the ZGB model in recent decades
has made its behavior and properties very well known, making it an excellent candidate for developing new approaches to IPT's. 

In this paper, we consider the application of the slowly-varying method to the first-order IPT of the ZGB model. We simulate the model using the standard ensemble and therefore the $CO$-pressure as the slowly-varying parameter. To characterize the dynamical behavior of the model, we employ a method based on the optimization of the coefficient of determination presented in \cite{daSilva2012} and subsequently used in several related problems \cite{daSilva2013a,daSilva2013b,daSilva2014,daSilva2015,daSilva2016,daSilva2020,Loscar2022}.
 
The remainder of this paper is organized as follows.
In section II, we provide the details of the model and explain the implementation of the DLP. We also define relevant quantities and describe the method of the coefficient of determination. In Section III, we present our numerical results of the DLP applied to the first-order IPT of the ZGB model. In section IV, we put our results into context in relation to previous works, and finally, we summarize the results and draw our conclusions.

\section{\label{sec:model} Model and Methods}

The ZGB model is a lattice-gas reaction model based upon the reaction of
carbon monoxide ($CO$) with oxygen ($O_2$) on a platinum catalyst surface.
In this model, the surface of the catalyst is simulated by a square lattice 
of side $L$ with periodic boundary conditions. In the simplest case, diffusion and desorption of reactants are ignored. The simulation proceeds according to the Langmuir-Hinshellwood mechanism, such that all molecules must adsorb before they can react. The temperature and pressure of the system are assumed to be such that the kinetics are adsorption limited, so that nearest-neighbor pairs of $CO$ and $O$ react and desorb immediately upon formation.  
We implement the following kinetic scheme
\begin{equation}
CO(g) + S \rightarrow CO(a) ,
\label{adco} 
\end{equation}
\begin{equation}
O_2(g) + 2S \rightarrow 2O(a) ,
\label{ado}
\end{equation}
\begin{equation}
CO(a) + O(a) \rightarrow CO_2(g)+2S ,
\label{reac}
\end{equation}

\noindent where $S$ refers to an empty lattice site, while ($a$) and ($g$) refer to the adsorbed phase and gas phase, respectively. The Monte Carlo simulation starts by
randomly selecting a lattice site. If the site is occupied, the trial finishes.
If the site is empty, with probability $p_{CO}$ a $CO$ molecule is adsorbed (Eq. (\ref{adco})) and with probability $1-p_{CO}$ a trial is made for the adsorption of the $O_2$ molecule. 
In the latter case, $O_2$ molecules require two adjacent (nearest-neighbor) empty sites since adsorption is dissociative. Then, a nearest-neighbor site is randomly chosen. If this site is empty, the $O_2$ molecule is adsorbed (Eq. (\ref{ado})); otherwise, the trial finishes. After the adsorption, the algorithm checks for adjacent neighbors occupied by different species. If a pair $CO$-$O$ is detected, they react immediately (Eq. (\ref{reac})), implying an infinite reaction rate compared to the adsorption rate. The only parameter in the model is $p_{CO}$, defined as the normalized rate (probability) that a molecule making an adsorption attempt is a $CO$. As usual, one Monte Carlo time step (MCS) involves $L^2$ trials, so that each site is visited on average once. The coverage of the species $X$ is given by $\theta_X=N_X/L^2$, where $N_X$ is the number of sites occupied by species $X$. We call this algorithm the {\it standard} (or the {\it constant rate}) ensemble. 

To account for the dynamic effects close to the first--order transition, we have implemented the DLP widely employed in equilibrium models \cite{tanos1,tanos2,tanos3,tanos4,castelnovo}. 
Firstly, we prepare a well-equilibrated configuration for a given pressure of $CO$ in the active window of the model near the transition point, such that $p_{CO}^{ini} \lesssim p_2$ where $p_2=0.525615$ is the transition pressure of the first-order IPT. This configuration is obtained by starting from an empty lattice and evolving the system 
with the standard ensemble at $p_{CO}=p_{CO}^{ini}$ during $\tau_{Relax}=5\times10^4$ MCS. After that, the pressure $p_{CO}$ is slowly varied as a linear function of the Monte Carlo time $t$ given by
\begin{equation}
p_{CO}= p_2(1+t/t_s)~,
\label{Eq.protocol}
\end{equation}
where $t_s$ is a parameter held constant during each simulation. In this method, $t_s$ is the characteristic time scale for the slowing variation experiment, which is introduced as a new independent parameter in the problem. 
From Eq. (\ref{Eq.protocol}), by replacing $p_{CO}=p_{CO}^{ini}$, one obtains the initial time $t=t_{ini}<0$ corresponding to the well-equilibrated initial state. Then, the pressure $p_{CO}$ is increased following Eq. (\ref{Eq.protocol}) until reaching the final pressure $p_f$. We choose $p_f$ in such a way that the system always ends in the poisoned state. 

The protocol is repeated by running many independent simulations and defining a \textit{dynamical} standard ensemble. In this ensemble, the averages of the observables are taken at fixed time $t$.
In this way, the average coverage of a species $X$ is given by
\begin{equation}
\langle \theta_{X}(t) \rangle=  \sum^{N_{run}}_{i=1} \theta_{X}^i(t)/N_{run} .
\label{Eq:theta}
\end{equation}
where $\theta_{X}^i(t)$ is the $X$-coverage at time $t$ in the i-th run, 
and $N_{run}$ is the number of independent runs. The generalized susceptibility, associated with the fluctuations of the $X$-coverage, is also given by
\begin{equation}
\chi_{X}(t) = L^2  \left[ \langle \theta_{X}(t)^2 \rangle - \langle \theta_{X}(t) \rangle^2 \right]~~,
\label{Eq:chi}
\end{equation}
where the symbols $\langle ~~  \rangle$ represent averages over independent runs at fixed time $t$, and $L$ is the linear size of the simulation lattice. We will consider the behavior of two order parameters: the coverage of $CO$ molecules ($\theta_{CO}$) and the density of empty (vacant) sites ($\theta_{vac}$).
Although the former is commonly used, some investigations have considered $\theta_{vac}$ as the order parameter \cite{Figueiredo2001,daSilva2016}. 

To apply the idea of finite-time effects to characterize a transition point, 
the appearance of typical power laws in the time scale $t_s$ must be considered. 
Therefore, one has to find (for example) extrapolated values with the best power law that fits the data. 
Here, we use a quantitative technique to find the best power law, employing a refinement technique proposed in \cite{daSilva2012} for a quantity called $\Phi$ as a function of $t_s$.
Let $(t_s^k,\Phi(t_s^k))$ be a set of data pairs to be fitted, where $t_s^k$ are different scale times with $k=1..N$. One defines the coefficient of determination $r$ \cite{daSilva2012} by
\begin{equation}
r=\frac{\sum\limits_{k=1}^{N}\left[~\overline{\ln \Phi }-a-b\ln t_s^k\right]^{2}}
{ \sum\limits_{k=1}^{N}\left[~\overline{\ln \Phi}-\ln \Phi(t_s^k)\right]^{2}}~,
\label{eq:coef_det}
\end{equation}
where $\overline{\cdot \cdot \cdot}$ stands for the average over the number of times $t_s$ included in the fit
\begin{equation*}
\overline{\ln \Phi}=\frac{1}{N}\sum\limits_{k=1}^{N}\ln \Phi(t_s^k)~\text{.}
\end{equation*}
In Eq. (\ref{eq:coef_det}) $a$ and $b$ are the linear coefficient and the slope, respectively, 
given by the linear fit of $\ln{\Phi}$ versus $\ln{t}$. The denominator of Eq. (\ref{eq:coef_det}) is the dispersion of the data around the average, whereas the numerator is the dispersion accounted for by the fit.
The resulting coefficient of determination is $r \leq 1$, with $r = 1$ holding for a fit that can explain all the variation, which is a perfect fit (see the appendix of reference \cite{daSilva2020} for more details).

\section{Numerical Results}

\begin{figure}
     \vspace{-0.5cm}
     \begin{tabular}{c}
       \includegraphics[width=6.0cm, bb=90 -20 550 550]{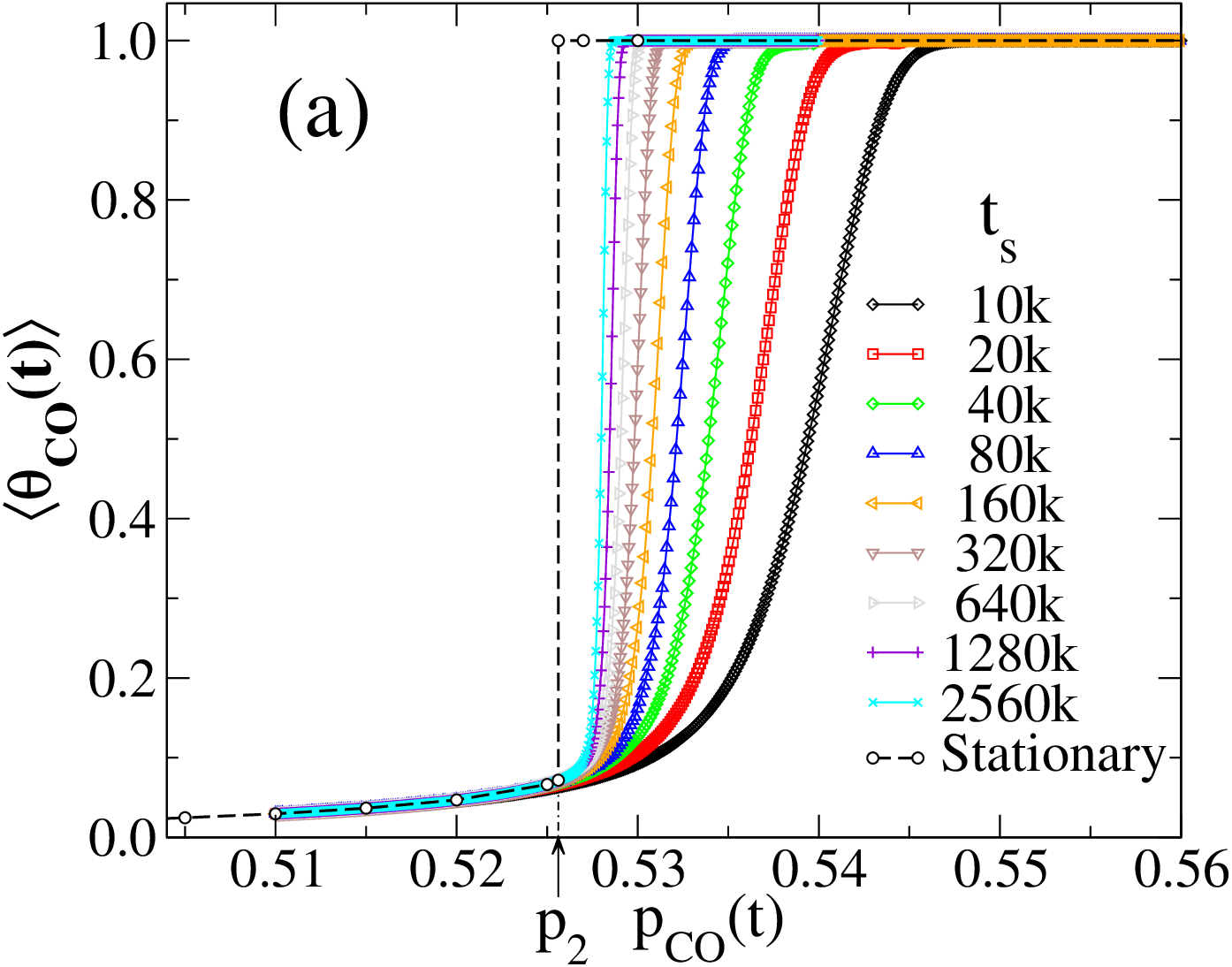}
       \vspace{0.3cm}
       \\ 
       \includegraphics[width=6.0cm, bb=90 -20 575 550]{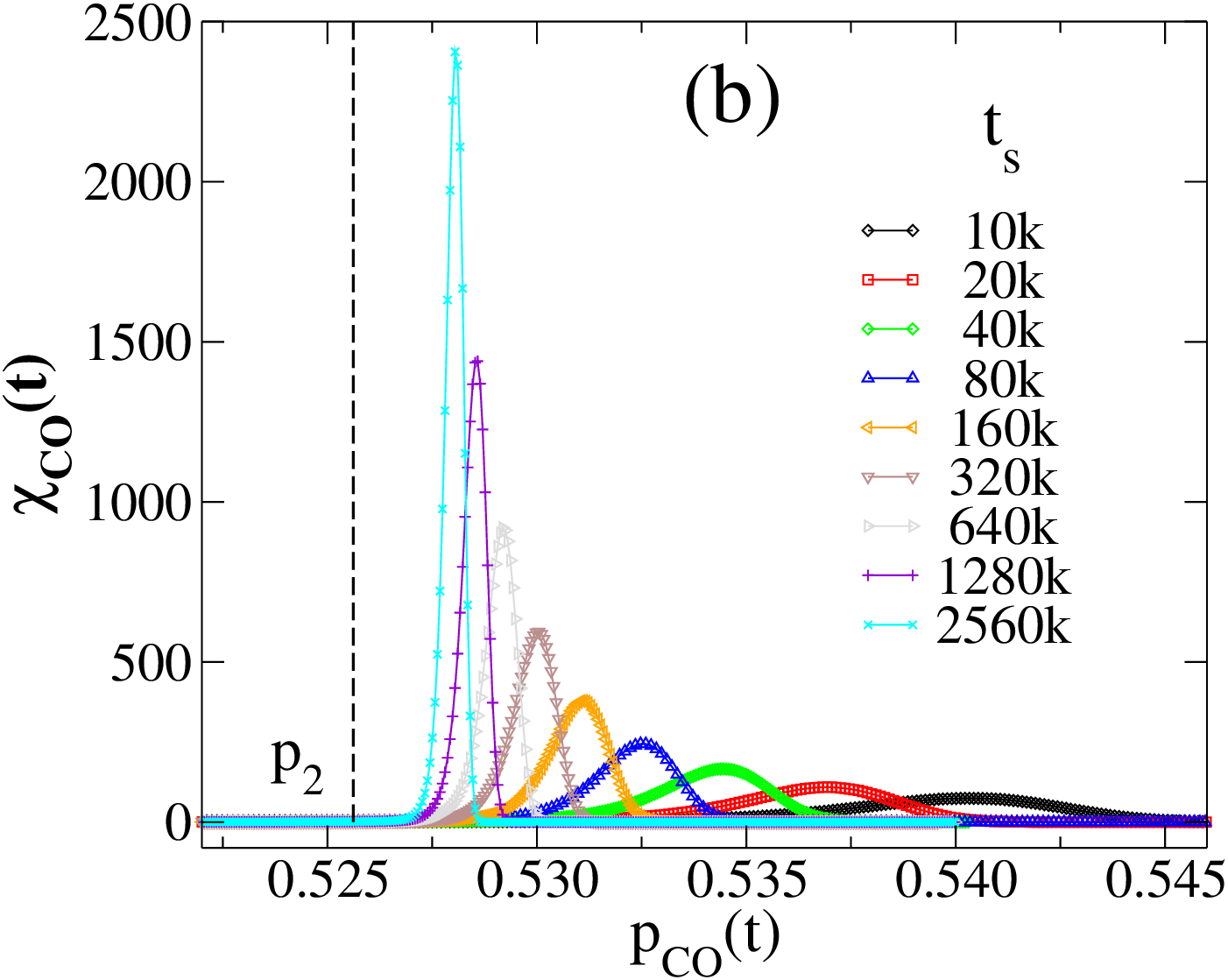}
       \vspace{0.3cm}
       \\
     \end{tabular}
\caption{Results of Monte Carlo simulation obtained with the DLP given by Eq. (\ref{Eq.protocol}) for a lattice of side $L=512$ and using different times $t_s$ as indicated. 
(a) Plots of $\langle \theta_{CO}(t) \rangle$ as a function of the control 
parameter $p_{CO}(t)$. Open black circles correspond to stationary results of Monte Carlo simulation obtained using the standard ensemble. 
(b) Plots of the susceptibility $\chi_{CO}(t)$ versus $p_{CO}(t)$. 
In both graph the vertical dashed lines correspond to the first-order IPT point $p_{2}=0.525615$. 
}
\label{fig.1}
\end{figure}

\begin{figure}
     \vspace{-0.5cm}
     \begin{tabular}{c}
       \includegraphics[width=6.0cm, bb=90 -20 550 550]{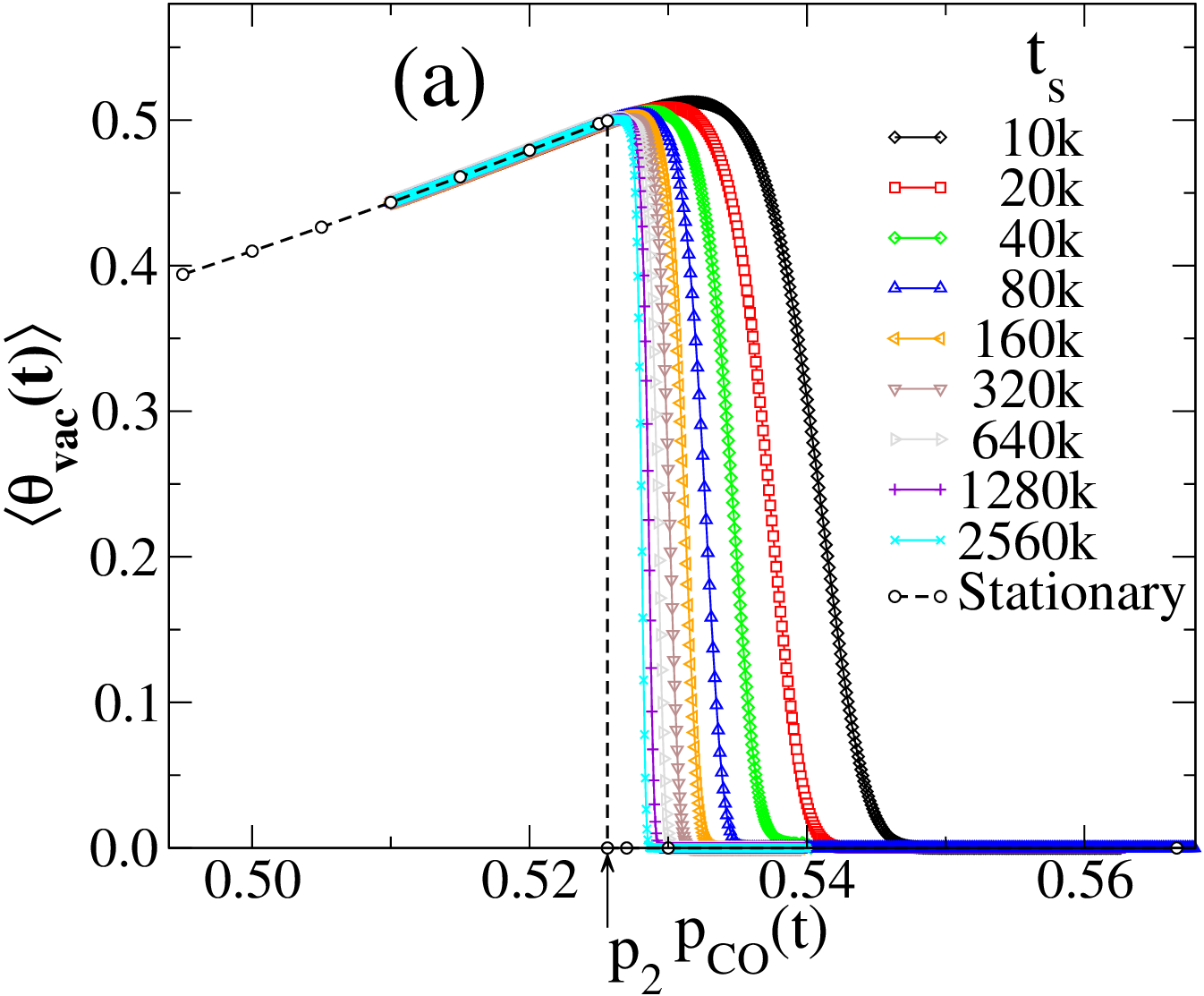}
       \vspace{0.3cm}
       \\ 
       \includegraphics[width=6.0cm, bb=90 -20 575 550]{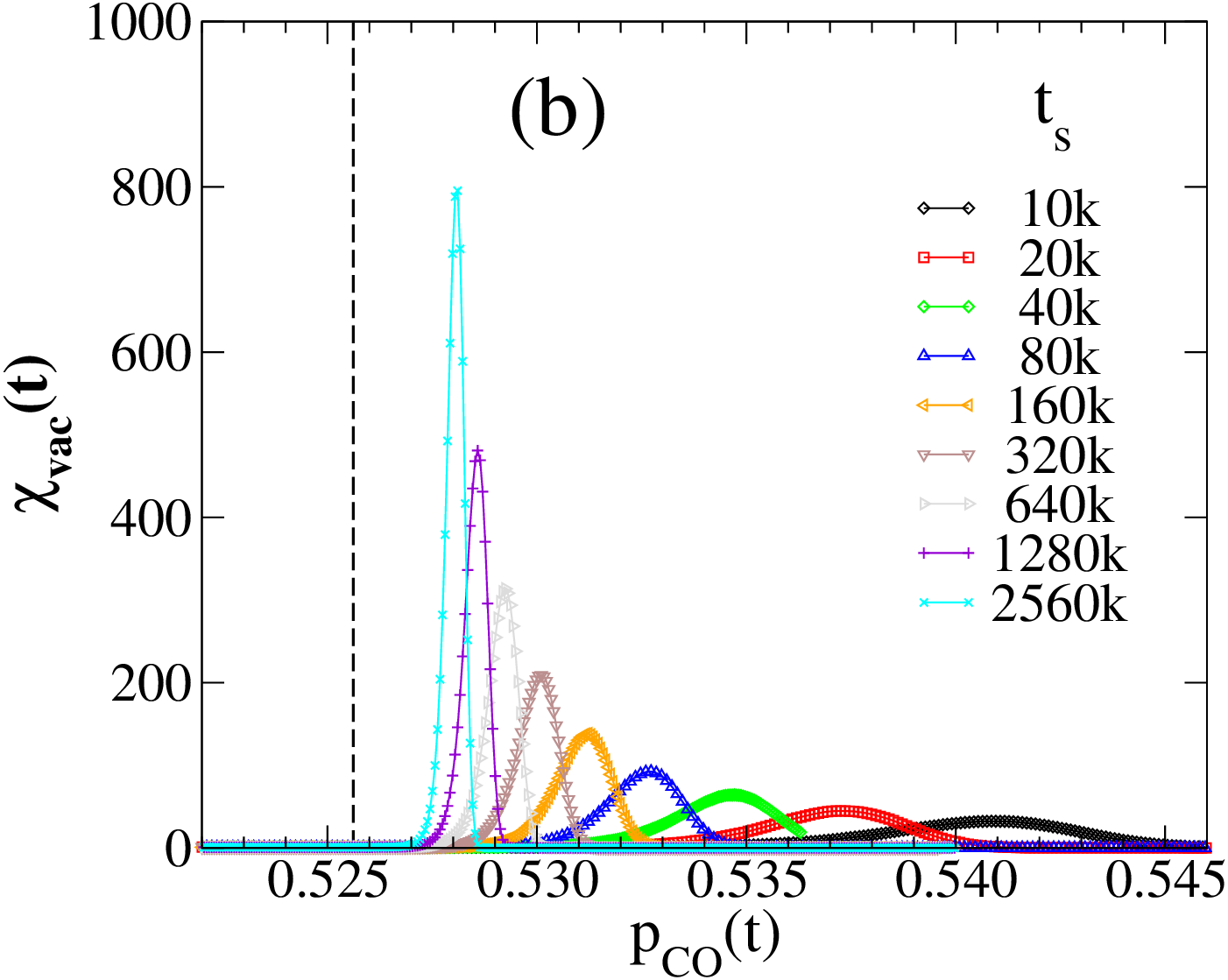}
       \vspace{0.3cm}
       \\
     \end{tabular}
\caption{Idem Fig. \ref{fig.1} for (a) the density of empty sites $\langle \theta_{vac}(t) \rangle$ and (b) susceptibility of empty sites $\chi_{vac}(t)$.
}
\label{fig.2}
\end{figure}
Firstly, we have simulated the standard ensemble and obtained the coverage of $CO$ 
and the density of empty sites, shown in Figs. \ref{fig.1}(a) and \ref{fig.2}(a), respectively, represented by open black circles and indicated as {\it stationary} results. These simulations start with an empty lattice, and then the system evolves with the standard ensemble at fixed pressure $p_{CO}$. We have used a square lattice of side $L=1024$ and $0.480\leq p_{CO}\leq p_2$ within the reactive window.
The first $10^5$ MCS are discarded, and the results are obtained from the average over the next $9\times10^5$ MCS. We have checked that these data are size-independent within the statistical error. 
Here, we will assume that the first-order IPT occurs at $p_{2}=0.525615$ from an active state with $CO$-coverage $\theta_2=0.0717$  to a poisoned state with $\theta_{CO}=1$. These values have been widely verified in the literature using several different simulation techniques \cite{zgb,evans1991, ziff1992,ccpnos,gradiente,monetti2001}. 

In order to study dynamic effects, we have performed extensive Monte Carlo simulations of the {\it dynamical} standard ensemble using lattices of different sizes $L$ and various values of time scales $t_s$. The time scales used range from $t_s=10$k to $t_s=2560$k. The number of samples used for each value of $t_s$ and $L$ was $N_{run}=10^4$. 
According to the DLP, we start the simulation with an initial well-equilibrated configuration for a given $p_{CO}^{in}$. This initial configuration is obtained by starting from an empty lattice and letting the system evolve during $5\times10^4$ MCS. 
Once the initial state is achieved, we apply the DLP given by Eq. (\ref{Eq.protocol}) with a fixed time scale $t_s$.

Figures \ref{fig.1}(a) and \ref{fig.2}(a) show the mean coverages of $CO$
and the mean density of empty sites, respectively, as functions of $p_{CO}(t)$.
These results were obtained using the DLP with the indicated 
time scales $t_s$ and a lattice of size $L=512$.
It is observed that the curves are very sensitive to the value 
of the time scale $t_s$, showing a rich dynamic behavior in the region $p_{CO} > p_2 $. The generalized susceptibilities $\chi_{CO}$ and $\chi_{vac}$ are shown versus $p_{CO}(t)$ in Figs. \ref{fig.1}(b) and \ref{fig.2}(b), respectively.
These functions develop a clear maximum that depends strongly on the time scale $t_s$.
It is worth mentioning, for this relatively large size $L=512$, that near the peaks of the susceptibilities $\chi_{CO}$ and $\chi_{vac}$, the density of empty sites $\theta_{vac}$ results $\sim 0.30$, while the probability of poisoning is null (not shown). Therefore, the response of the system to perturbations with different time scales $t_s$ is maximum for active states and involves non-trivial correlations and fluctuations. We will study this response in the following paragraphs.

\begin{figure}
\includegraphics[width=8cm]{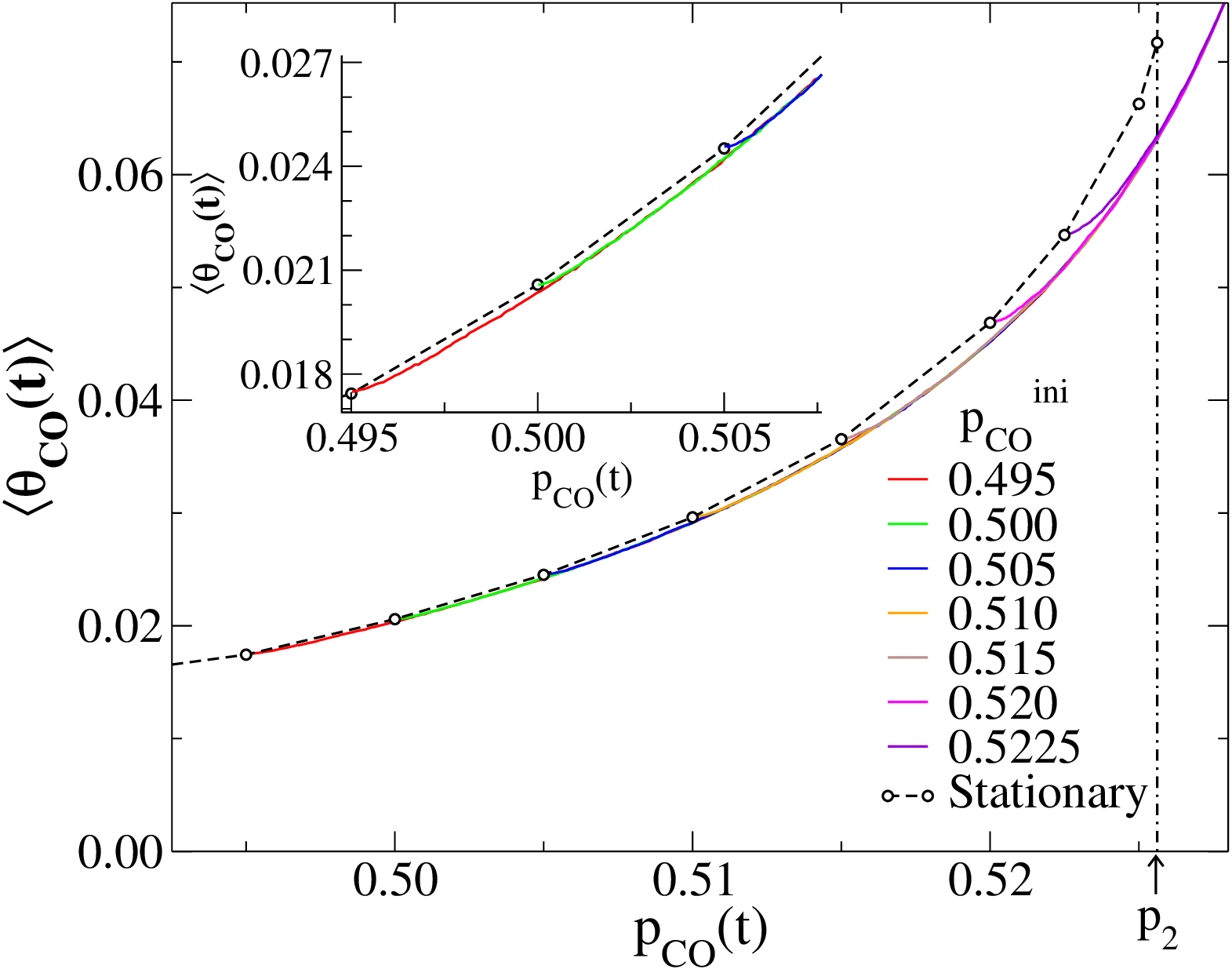}
\caption{Plots of $\langle\theta_{CO}(t)\rangle$ as a function of the control parameter $p_{CO}(t)$. Color lines are results of Monte Carlo simulations obtained with the DLP (\ref{Eq.protocol}) for time scale $t_s=10k$. Different curves correspond to indicated initial pressure $p_{CO}^{ini}$. Open circles correspond to stationary results obtained by using the standard ensemble (dashed line is used only to guide the eyes). Vertical dashed line corresponds to the first-order IPT point $p_{2}=0.525615$.}
\label{fig.cini}
\end{figure}

Figure \ref{fig.cini} shows the results of the dynamical standard ensemble starting with different initial pressures $p_{CO}^{ini}$, using a fixed time scale $t_s=10$k, and a lattice of size $L=128$. 
For comparison, stationary data obtained with the standard ensemble are shown.
We can observe, after a short period of time, that the evolution of the coverage 
always converges to a unique curve $\theta_{CO}(p_{CO},t_s)$ which depends only on the time scale $t_s$. As will be demonstrated below, these results are also independent of the size of the system.

In the region where $p_{CO}\le p_2$, the curve $\theta_{CO}(p_{CO},t_s)$ is clearly below the stationary results. 
These data depend only on the finite time scale $t_s$ and can be considered 
{\it finite time} effects (analogous to finite-size effects). Therefore,
they can be extrapolated to the limit $t_s\rightarrow\infty$. 
We expect that these extrapolated results coincide with the stationary ones. 
The main conclusion obtained from Fig. \ref{fig.cini} is that when running the DLP, there is no dependency on the initial pressure $p_{CO}^{ini}$.

\begin{figure}
\includegraphics[width=14cm]{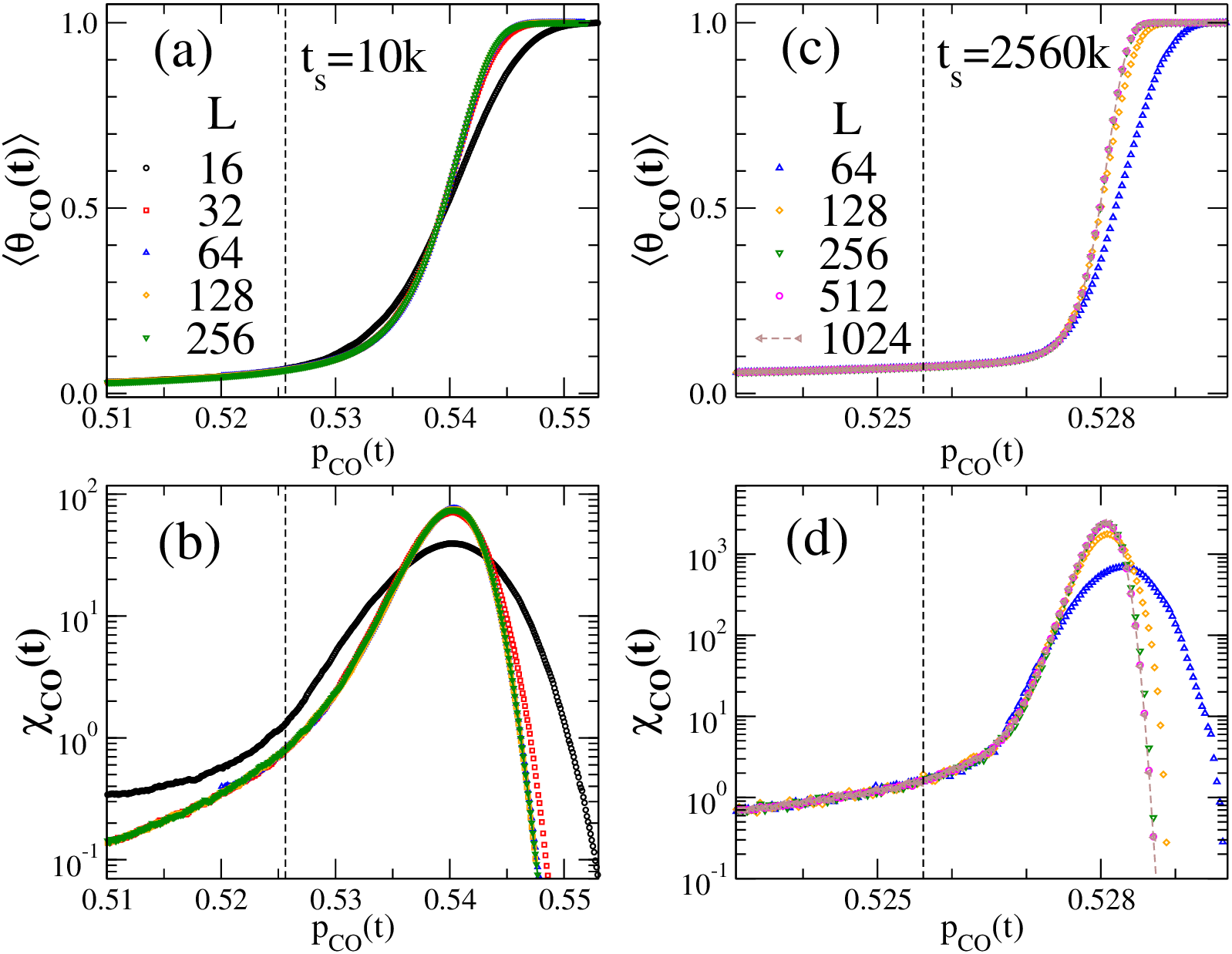}
\caption{
Plots of (a) coverage $\theta_{CO}(t)$ and (b) the susceptibility $\chi_{CO}(t)$ versus the control parameter $p_{CO}(t)$. The data were obtained with the DLP (\ref{Eq.protocol}) for $t_s=10$k and different sizes $L$ as indicated. Size independence is observed for $L\geq 64$. 
Plots of (c) coverage $\theta_{CO}(t)$ and (d) susceptibility $\chi_{CO}(t)$ versus the control parameter $p_{CO}(t)$. The data were obtained with the DLP (\ref{Eq.protocol}) for $t_s=2560$k and different sizes $L$ as indicated. Size independence is observed for $L\geq 256$. Vertical dashed lines indicate the first order IPT point $p_{2}=0.525615$.
}
\label{fig.satur}
\end{figure}
In order to gain insight into the behavior at higher pressures, it is useful 
to analyze results for $p_{CO}>p_2$ fixing the time $t_s$.
Figure \ref{fig.satur} shows the mean coverage $\langle\theta_{CO}(t)\rangle$ 
and the susceptibility $\chi_{CO}(t)$ of the $CO$-species, given by Eqs. (\ref{Eq:theta}) and (\ref{Eq:chi}) 
respectively, versus the pressure $p_{CO}(t)$ given by Eq. (\ref{Eq.protocol}).

The results obtained for the minimum time scale used in this work $t_s=10$k are 
shown in Figs. \ref{fig.satur}(a) and \ref{fig.satur}(b). From these figures,
for small systems ($L<64$), we can observe that our results depend on the system size. In this regime, a detailed knowledge of the poisoning time is necessary to  
introduce the system size $L$ in a complete description of the behavior of the system. However, for larger sizes ($64\leq L$), there is no dependence on size $L$.
This size-independent behavior is observed for both coverage 
$\langle\theta_{CO}(t)\rangle$ (Fig. \ref{fig.satur}(a)) and fluctuations $\chi_{CO}(t)$ (Fig. \ref{fig.satur}(b)).
We will call the minimum size from which we observe size independence
the saturation length $L_{\texttt{sat}}$. Therefore, for $t_s=10$k we have $L_{\texttt{sat}}\approx64$. The same analysis can be performed 
with data shown in Figures \ref{fig.satur}(c) and \ref{fig.satur}(d), obtained from simulations for the maximum time scale studied in this work $t_s=2560$k. 
In this case, the data show size independence for $L_{\texttt{sat}}\approx 256$. Thus, the saturation length $L_{\texttt{sat}}$ is an increasing function of the time scale $t_s$. 

In the following, we will study only the finite-time effects. In order to do that,
we eliminate the finite-size effects by taking the thermodynamic limit $L\rightarrow\infty$ for each $t_s$. In practice, this regime is reached when the condition $L>L_{\texttt{sat}}$ is satisfied.
The results shown in Figs. \ref{fig.1} and \ref{fig.2} were obtained with samples whose sizes saturate in this sense. Therefore, we expect that all the effects shown in Figs. \ref{fig.1} and \ref{fig.2} are due to the finite time scale $t_s$. We will refer to these effects as {\it finite-time} effects. 

Interestingly, it should be noted that these finite-time effects would produce scaling laws, analogous to finite-size scaling, in the limit of large times $t_s\rightarrow\infty$. Then, for a quantity $\Phi_X$  measured with the dynamical standard ensemble, we have schematically that the order in which we take the limits is 
\begin{equation}
\lim_{t_s\to\infty} \left[\lim_{L\to\infty}  \Phi_{X}(p_{CO},L,t_s) \right]
\label{Eq:limits}
\end{equation}
where $\Phi$ is $\theta$ or $\chi$, and $X$ is $CO$ or $vac$. We emphasize that our
DLP is applied to the region where the first-order phase transition
takes place; therefore, the order of the limits will be significant.
 \begin{figure}
   \begin{center}
     \begin{tabular}{c}
       \includegraphics[width=5.0cm, bb=70 -20 550 550]{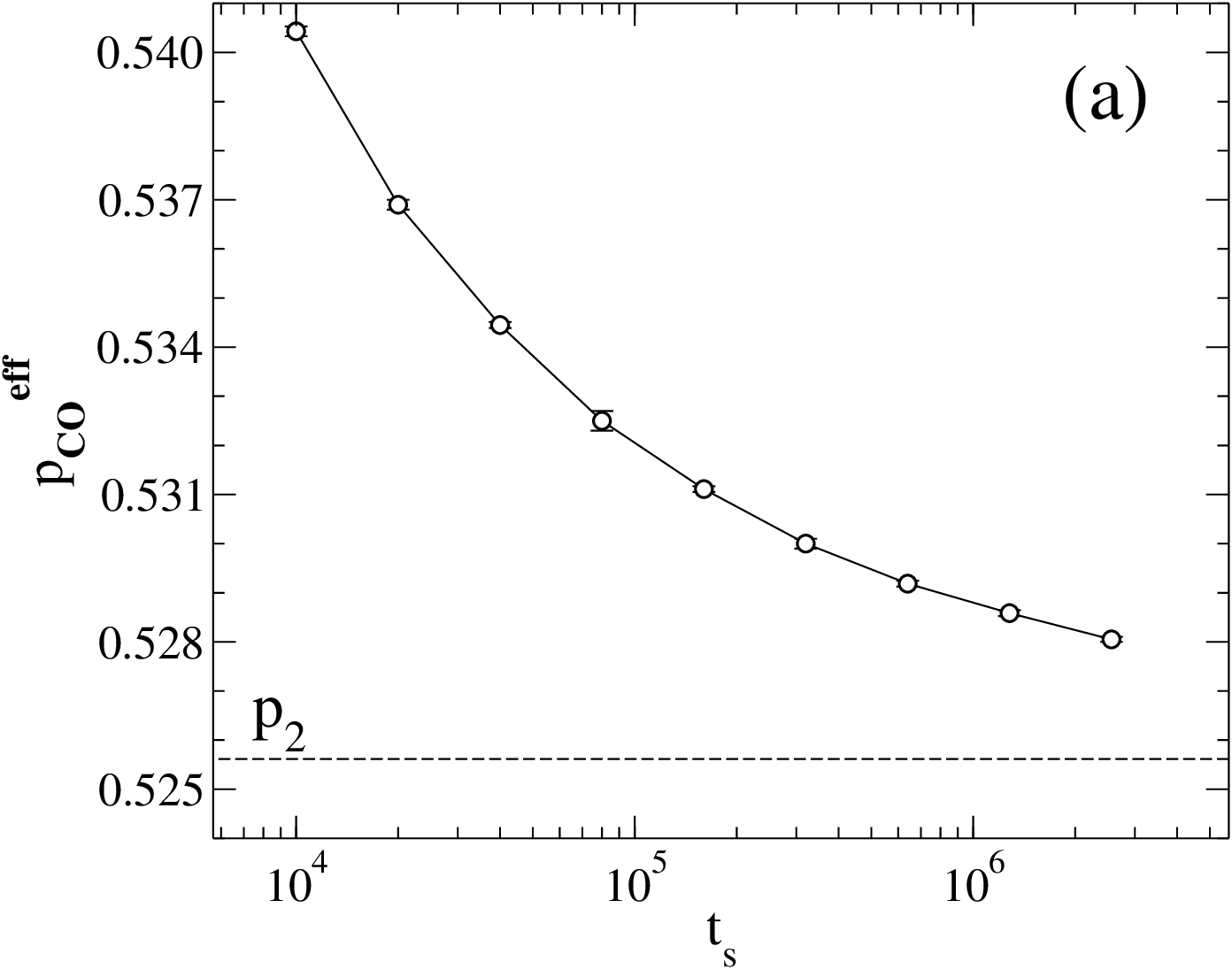}
       \\ 
       \includegraphics[width=5.0cm, bb=70 -20 550 550]{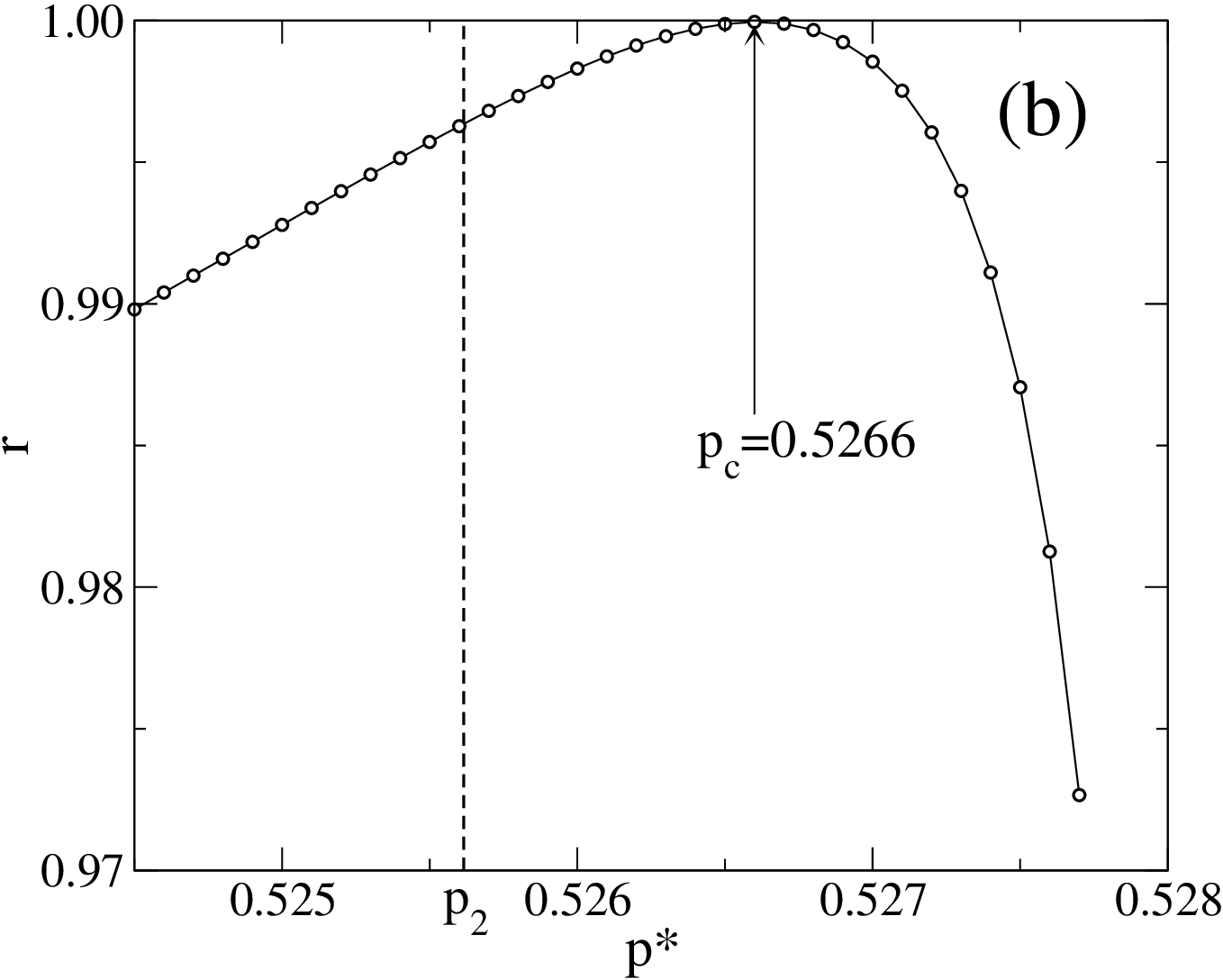}
       \\
       \includegraphics[width=5.0cm, bb=70 -20 550 550]{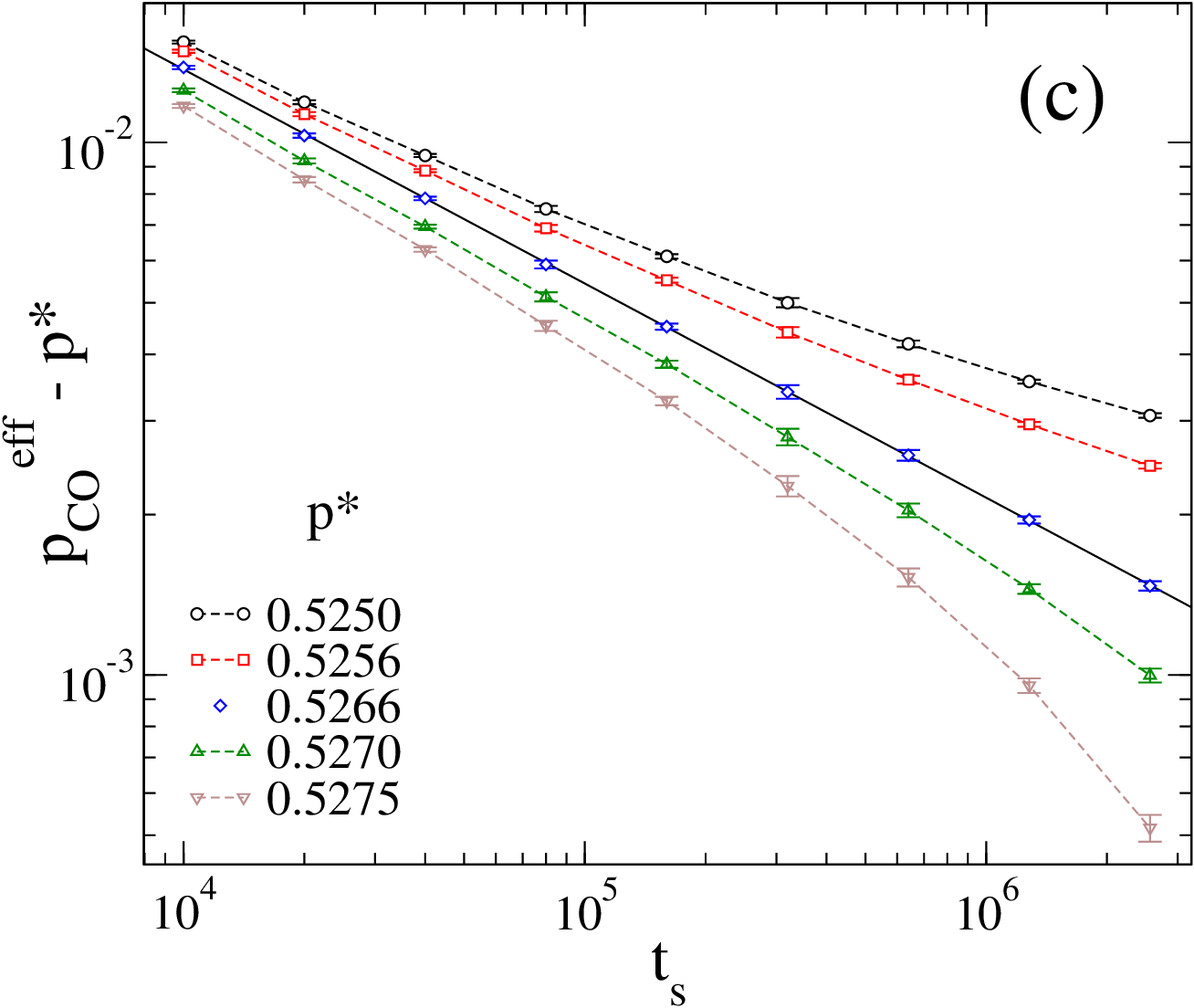}
     \end{tabular}
     \end{center}
     \caption{Determination of the transition pressure $p_c$ using the $CO$-coverage.
     (a) Effective transition pressure $p^{eff}_{CO}$ corresponding to the maximum susceptibility of $CO$, showed in Fig. \ref{fig.1}, as function of the time scale $t_s$. Horizontal line corresponds to the first-order IPT point $p_2$. 
     (b) Plot of the coefficient of determination $r$ calculated for several $CO$-pressures $p^*$. The best fit is obtained for $p^*=0.5266$ as indicated.
     (c) Log-log plots of the difference $p^{eff}_{CO}-p^*$ versus $t_s$. The continuous line is the best fit obtained for $p^*=0.5266$.}
     \label{fig.PeffCO}
 \end{figure}

\begin{figure}
   \begin{center}
     \begin{tabular}{c}
       \includegraphics[width=5.0cm, bb=70 -20 550 550]{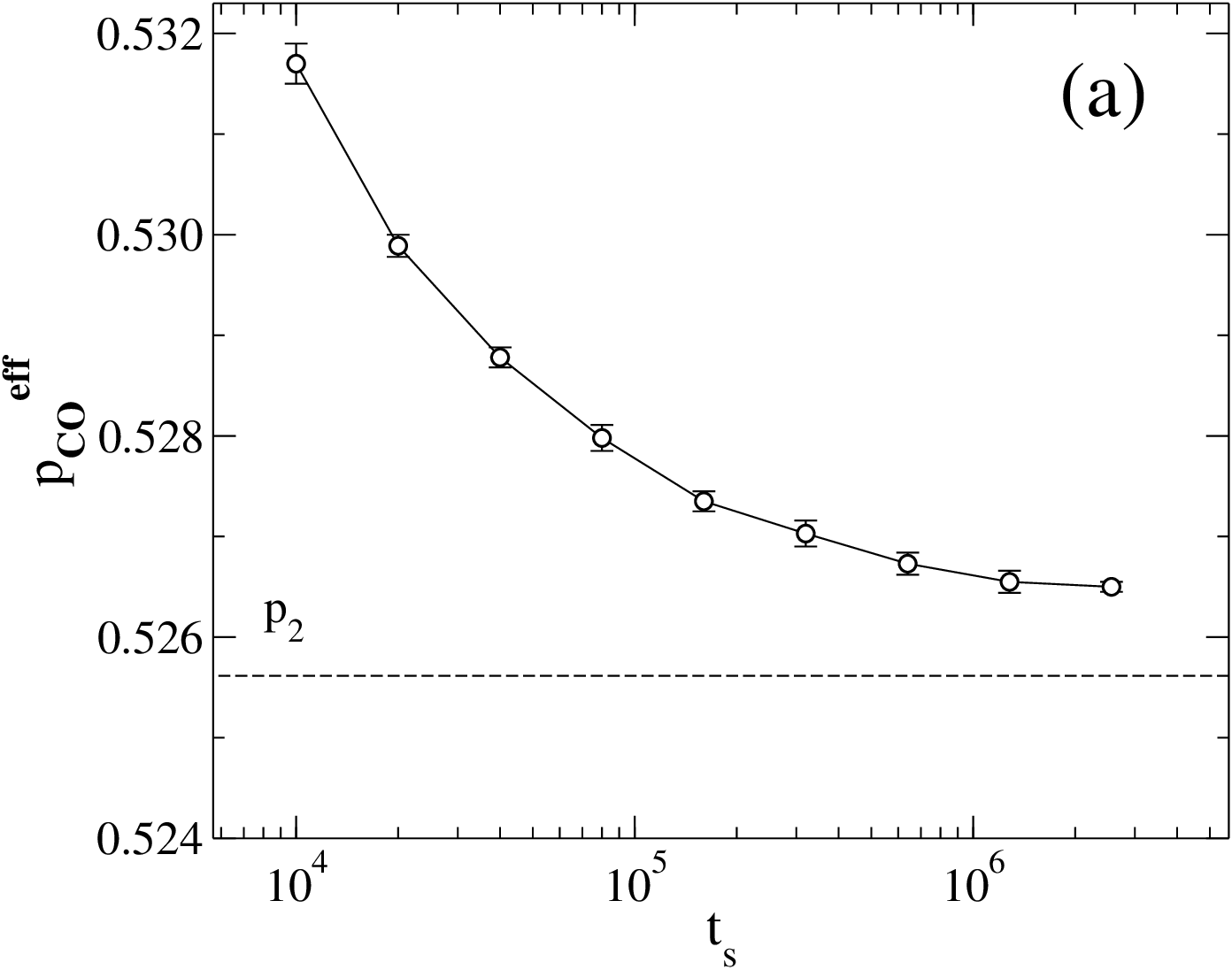}
       \\ 
       \includegraphics[width=5.0cm, bb=70 -20 550 550]{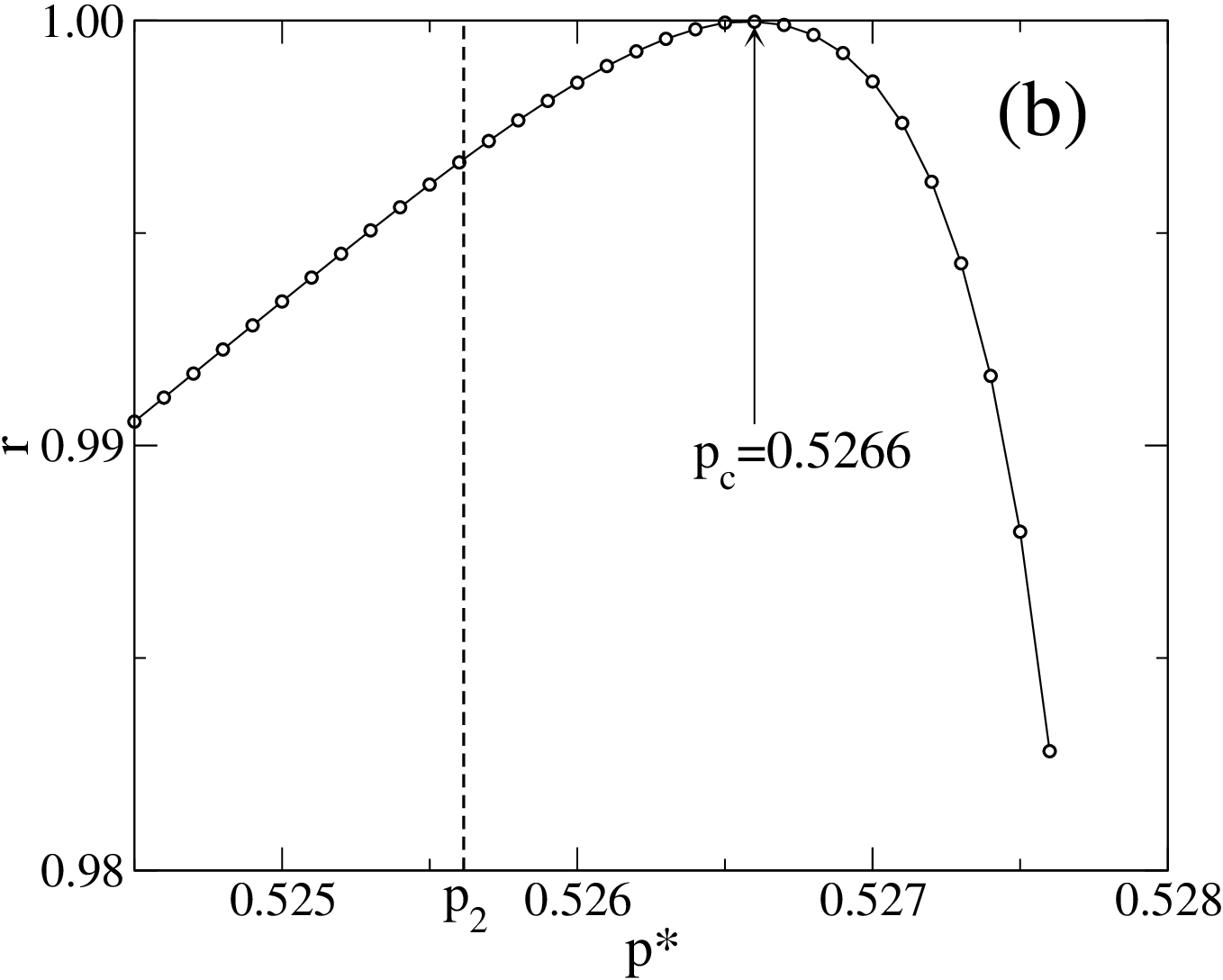}
       \\
       \includegraphics[width=5.0cm, bb=70 -20 550 550]{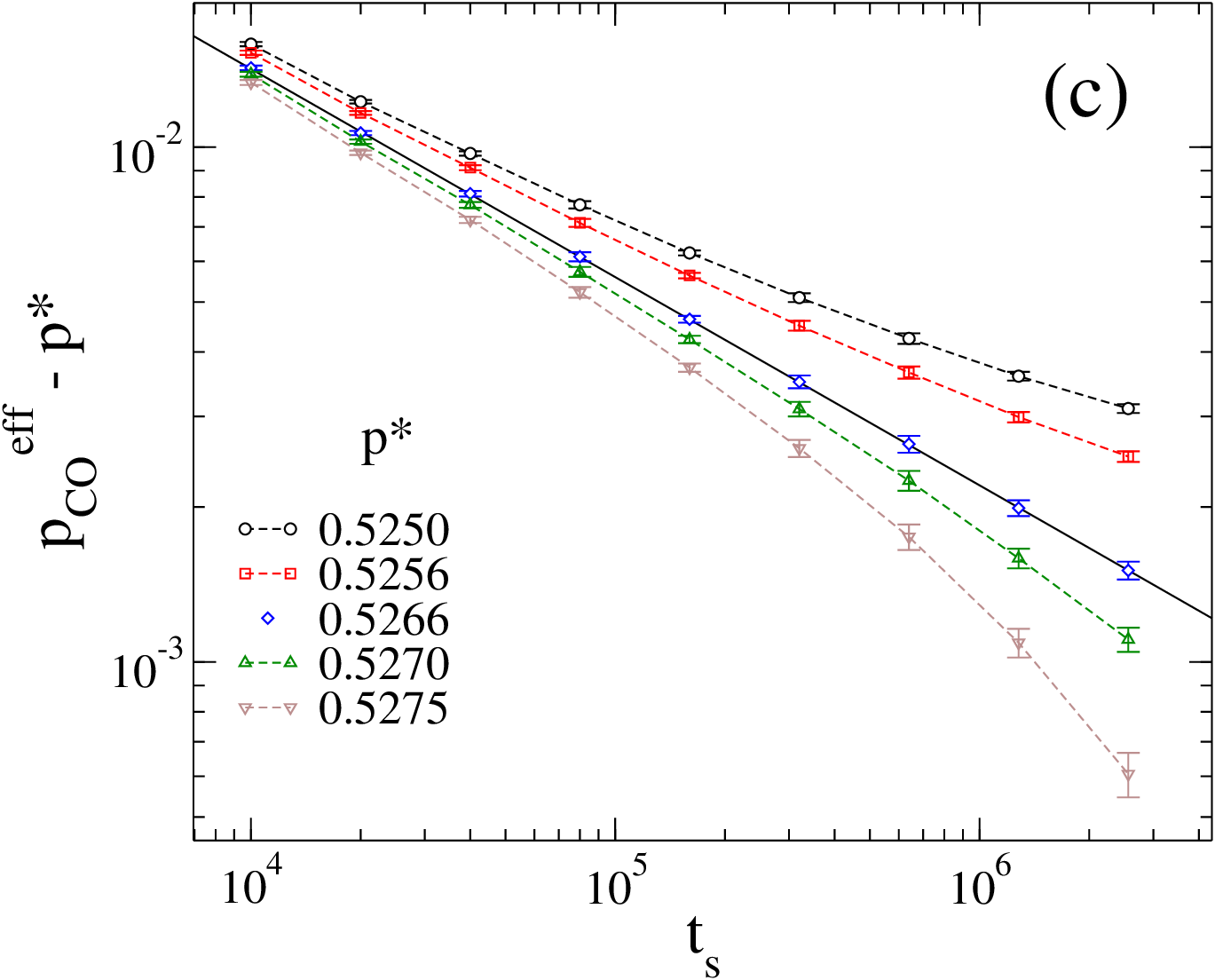}
     \end{tabular}
   \end{center}
   \caption{Idem Fig. \ref{fig.PeffCO} for effective pressure obtained with vacant sites.}
   \label{fig.PeffVac}
\end{figure}
The results shown in Figs. \ref{fig.1} and \ref{fig.2} depend only on the new free parameter $t_s$, the time scale introduced in Eq. (\ref{Eq.protocol}). 
We observe that the susceptibilities of the $CO$-species and the empty sites
(Figures \ref{fig.1}(b) and \ref{fig.1}(b)) develop a clear peak depending on each time $t_s$. In addition, the height of the peaks increases strongly with $t_s$. 
On the other hand, the coverages $\theta_{CO}$ and $\theta_{vac}$ shown in Figs. \ref{fig.1}(a) and \ref{fig.2}(a), respectively, have a sigmoid shape around the maximum of susceptibilities. These rounded discontinuities are more abrupt for larger times $t_s$. Both curves, susceptibilities and coverages, are shifted towards lower values as the time $t_s$ increases.
This behavior resembles the finite-size scaling for phase transitions \cite{BinderLibro}, with 
the new parameter $t_s$ as the scaling variable instead of the linear size $L$. We remark that finite-size laws are expressed in terms of spatial dimensions $L$, and they are valid in the limit of the time scale going to infinity, that is, for the stationary regime. 
Analogously, here we propose that finite-time effects be expressed in terms of the {\it dynamic dimension} $t_s$ with the size scale going to infinity. 
In this way, following this idea, we can interpret the pressure corresponding to 
the maximum susceptibilities $p_{CO}^{eff}(t_s)$ in Figs. \ref{fig.1}(b) and 
\ref{fig.2}(b) as an effective value (for finite $t_s$) of {\it an apparent} phase transition.

Figures \ref{fig.PeffCO}(a) and \ref{fig.PeffVac}(a) show the effective value of the pressure $p_{CO}^{eff}$ versus $t_s$, calculated with the peaks of the $CO$ susceptibility ($\chi_{CO}$) and with the peaks of the susceptibility of vacant sites ($\chi_{vac}$), respectively. 
Then, the critical pressure of this apparent transition can be extrapolated using a method based on standard scaling laws: 
\begin{equation}
p_{CO}^{eff}(t_s)= p_{CO}^{crit}(t_s\rightarrow\infty) + A t_s^{-1/\nu_X}
\label{Eq:pcocrit}
\end{equation}
where $A$ is a constant, $p_c\equiv p_{CO}^{crit}(t_s\rightarrow\infty)$ is the extrapolated transition pressure, and $\nu_X$ is a characteristic time-correlation exponent.

In order to determine $p_c$ and $\nu_X$, we will apply the $r$-coefficient method 
described in section \ref{sec:model}. Firstly, we consider the quantity 
$p_{CO}^{eff}(t_s)-p^*$ 
versus $t_s$ for different arbitrary values of $p^*$. 
Then, we calculate the coefficient of determination $r$, given by Eq. (\ref{eq:coef_det}), for each value of $p^*$ and look for the maximum value of $r$ that corresponds to the best power law. Figs. \ref{fig.PeffCO}(b) and \ref{fig.PeffVac}(b) show the coefficient $r$ versus $p^*$ calculated with the effective pressure shown in Figs. \ref{fig.PeffCO}(a) and \ref{fig.PeffVac}(a), respectively. The maxima $r\approx 1$ are indicated with an arrow and correspond to a transition pressure $p_c=0.5266\pm0.0001$ in both cases. 

Figures \ref{fig.PeffCO}(c) and \ref{fig.PeffVac}(c) show on log-log scales the differences $p_{CO}^{eff}(t_s)-p^*$ versus $t_s$ for various values of $p^*$. The straight lines are a power law fit for $p_c=0.5266$. The fitted exponents are $1/\nu_{CO}=0.40 \pm 0.01$ (Fig. \ref{fig.PeffCO}(c)) and $1/\nu_{vac}=0.41 \pm 0.01$ (Fig. \ref{fig.PeffVac}(c)). As can be seen in both figures,
other values of $p^*$ show deviations from the power-law in good agreement with the r-coefficient shown in Figs. \ref{fig.PeffCO}(b) and \ref{fig.PeffVac}(b).

It is worth mentioning that the extrapolated transition pressure $p_c=0.5266$ is quite close to the pressure corresponding to the first-order IPT point $p_2=0.525615$. 
However, as shown in Figs. \ref{fig.PeffCO}(b) and \ref{fig.PeffVac}(b), the transition point $p_2$ (indicated by the vertical dashed line) is significantly lower than the pressure $p_c$ corresponding to the maximum $r$. We also note that near this apparent transition, the surface of the catalyst is always in an active state, {\it i.e.} the probability of being poisoned is zero. 
It is a consequence of the fact that the coverages of $CO$ and vacant sites
(shown in Figs. \ref{fig.1} and \ref{fig.2}) are $\theta_{CO}\approx 0.60$ and 
$\theta_{vac}\approx 0.30$, respectively, at the peak of susceptibilities.

 \begin{figure}
   \begin{center}
     \begin{tabular}{c}
       \includegraphics[width=5.0cm, bb=70 -20 550 550]{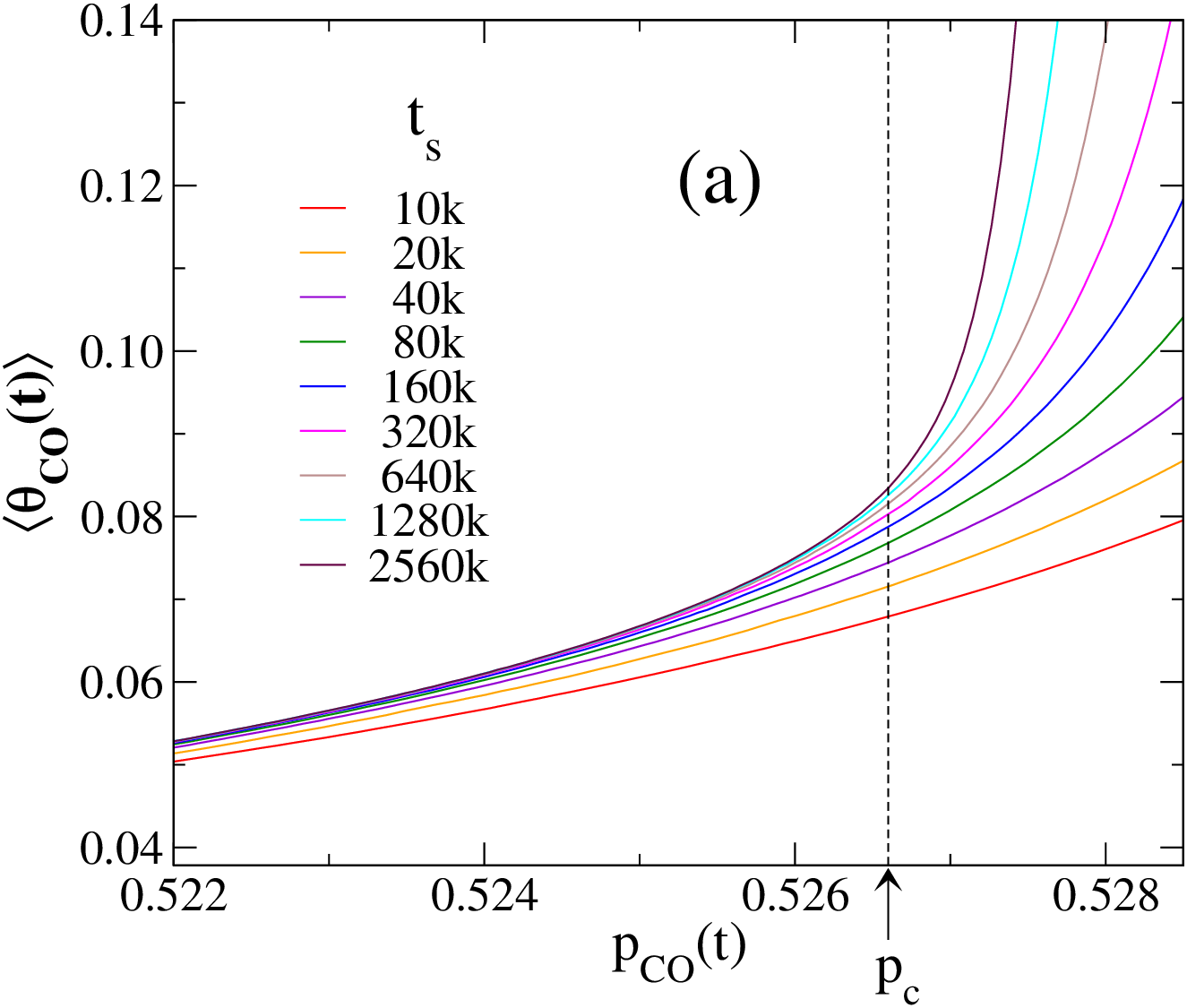}
       \\ 
       \includegraphics[width=5.0cm, bb=70 -20 550 550]{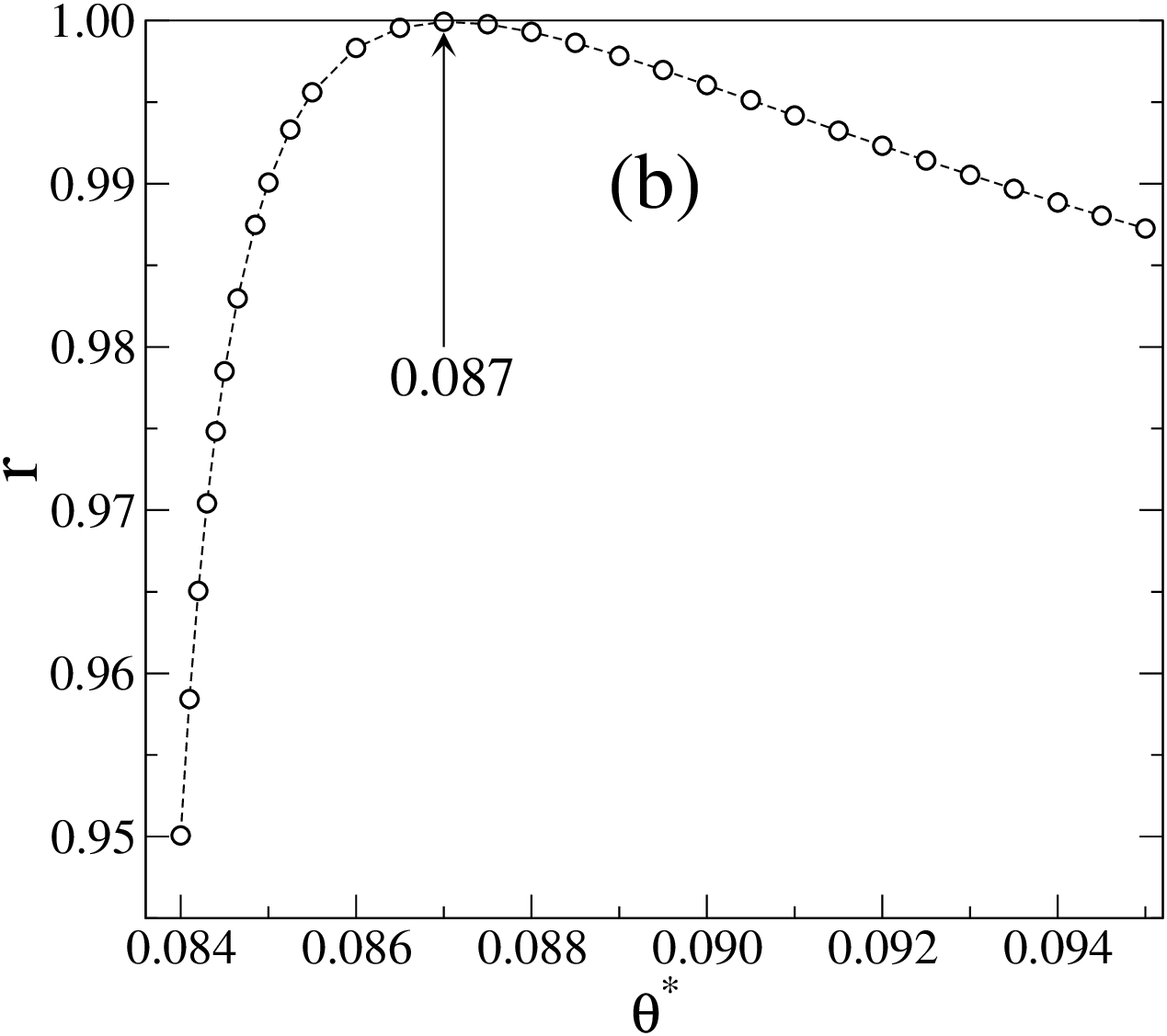}
       \\
       \includegraphics[width=5.0cm, bb=70 -20 550 550]{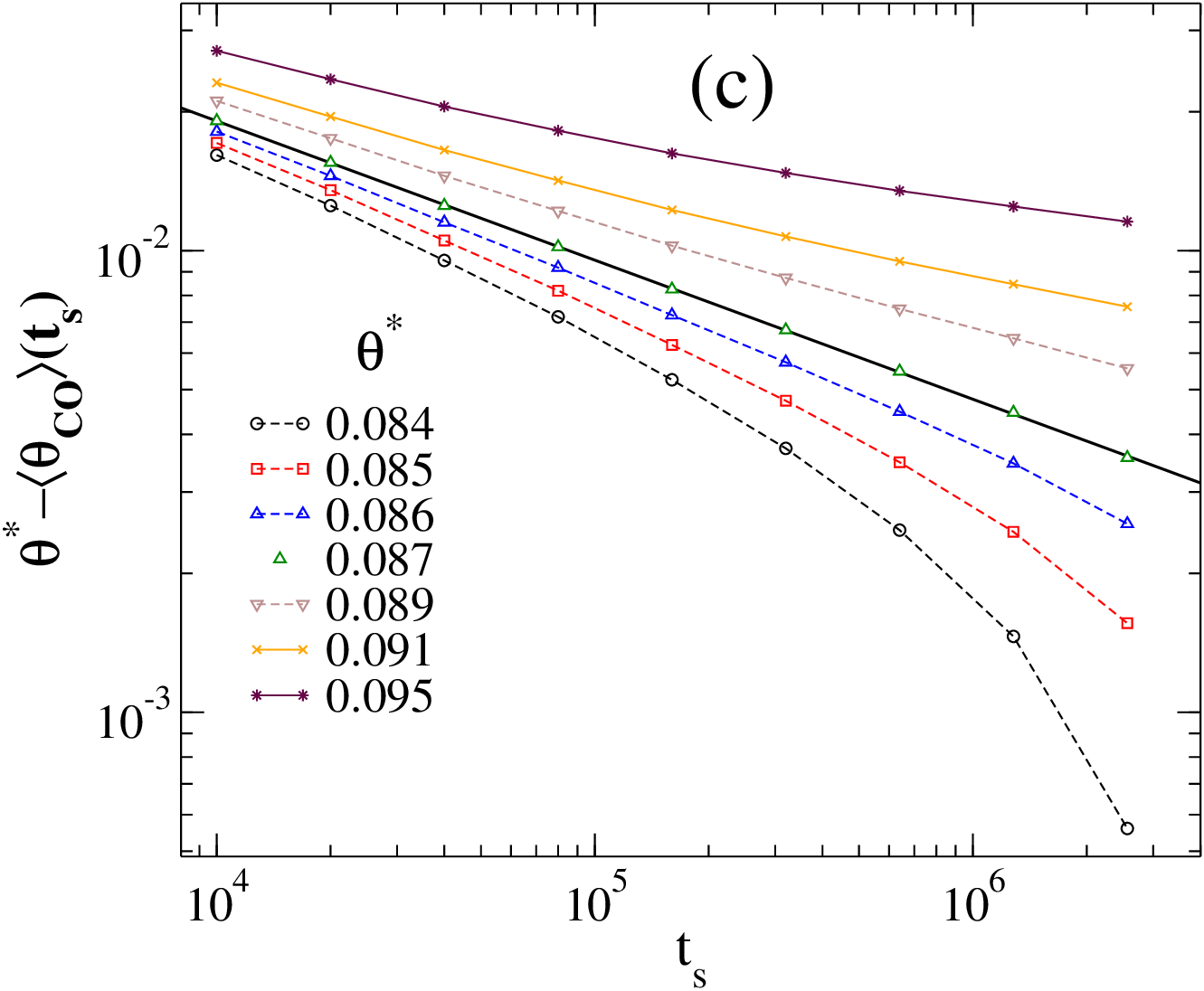}
     \end{tabular}
     \caption{(a) Plots of $\langle\theta_{CO}(t)\rangle$ as function of the control parameter $p_{CO}(t)$. Color lines are results of Monte Carlo simulation obtained with the DLP given by Eq. (\ref{Eq.protocol}) for different time scales $t_s$ as indicated. Vertical dashed line corresponds to the apparent transition point obtained in Figs. \ref{fig.PeffCO} and \ref{fig.PeffVac}: $p_c=0.5266$.
     (b) The coefficient of determination calculated for several arbitrary coverages $\theta^*$
     with the scaling law given in Eq. (\ref{Eq.powerlawtheta}).
     The best fit is obtained for $\theta^*=0.0870$ as indicated.
     (c) Log-log plots of the $\theta^*-\theta_{CO}(t_s,p_c)$ versus $t_s$. 
     The continuous line shows the best fit obtained for $\theta^*=0.0870$.
     Symbols are greater than the statistical errors.
     }
     \label{fig.extrapol}
   \end{center}
 \end{figure}
Once the $CO$-pressure $p_c$ of the apparent transition is known by means of the extrapolation given in Eq. (\ref{Eq:pcocrit}), the behavior of other quantities can be studied. One example is the characteristic behavior of the $CO$-coverage. 
We consider the coverage curves $\theta_{CO}(t)$ as a function of the driven pressure $p_{CO}(t)$ corresponding to different time scales $t_s$ shown in Fig. \ref{fig.extrapol}(a).
We remark that these data are independent of the size of the system $L$ and the initial states and can be represented by a function of the form $\theta_{CO}(p_{CO},t_s)$.
For finite scale time $t_s$, the coverage of $CO$ associated with the transition 
is given by the value $\theta_{CO}(p_c,t_s)$, obtained as the intersection of the
vertical line $p_{CO}=p_c$ with the dynamical curves $\theta_{CO}(p_{CO},t_s)$, as shown in Fig. \ref{fig.extrapol}(a).
Following the idea of finite-time scaling, we will assume that this coverage, for finite time $t_s$, tends to an extrapolated value by means of a typical scaling law
\begin{equation}\label{Eq.powerlawtheta}
\phi \propto  t_s^{-\Delta}
\end{equation}
where $\phi$ is $ \phi\equiv \theta_{c} - \theta_{CO}(p_c,t_s)$, and $\Delta$ is the characteristic exponent. In order to determine $\theta_c$ and $\Delta$, 
we will apply the $r$-coefficient method again. We consider the quantity $\theta^*-\theta_{CO}(p_c,t_s)$ versus $t_s$ for several arbitrary values 
of $\theta^*$. Then, we calculate the coefficient $r$ given by Eq. (\ref{eq:coef_det}) for each value of $\theta^*$ and look for the best power law.  
As shown in Fig. \ref{fig.extrapol}(b), we obtain the maximum coefficient $r$ ($r\approx1$) for $\theta^*=0.087$, indicated by an arrow, which determines the best power law. 
Figure \ref{fig.extrapol}(c) shows $\left[\theta^*-\theta_{CO}(p_c,t_s)\right]$ versus $t_s$
on a log-log scale for different values of $\theta^*$, as indicated. The continuous line is the best fit for the power law given by Eq. (\ref{Eq.powerlawtheta}), obtained for $\theta^*=\theta_c=0.087$, which produces the fitted exponent $\Delta=0.32\pm0.02$. We note that this transition $CO$-coverage 
$\theta_c=0.0870\pm0.0005$ is quite different from the value corresponding to the gas phase in coexistence at the first-order transition $\theta_2=0.0717$ \cite{ccpnos,ccpnos2}.
\begin{figure}
   \begin{center}
     \vspace{-0.5cm}
     \begin{tabular}{c}
       \includegraphics[width=5.50cm, bb=70 -20 550 550]{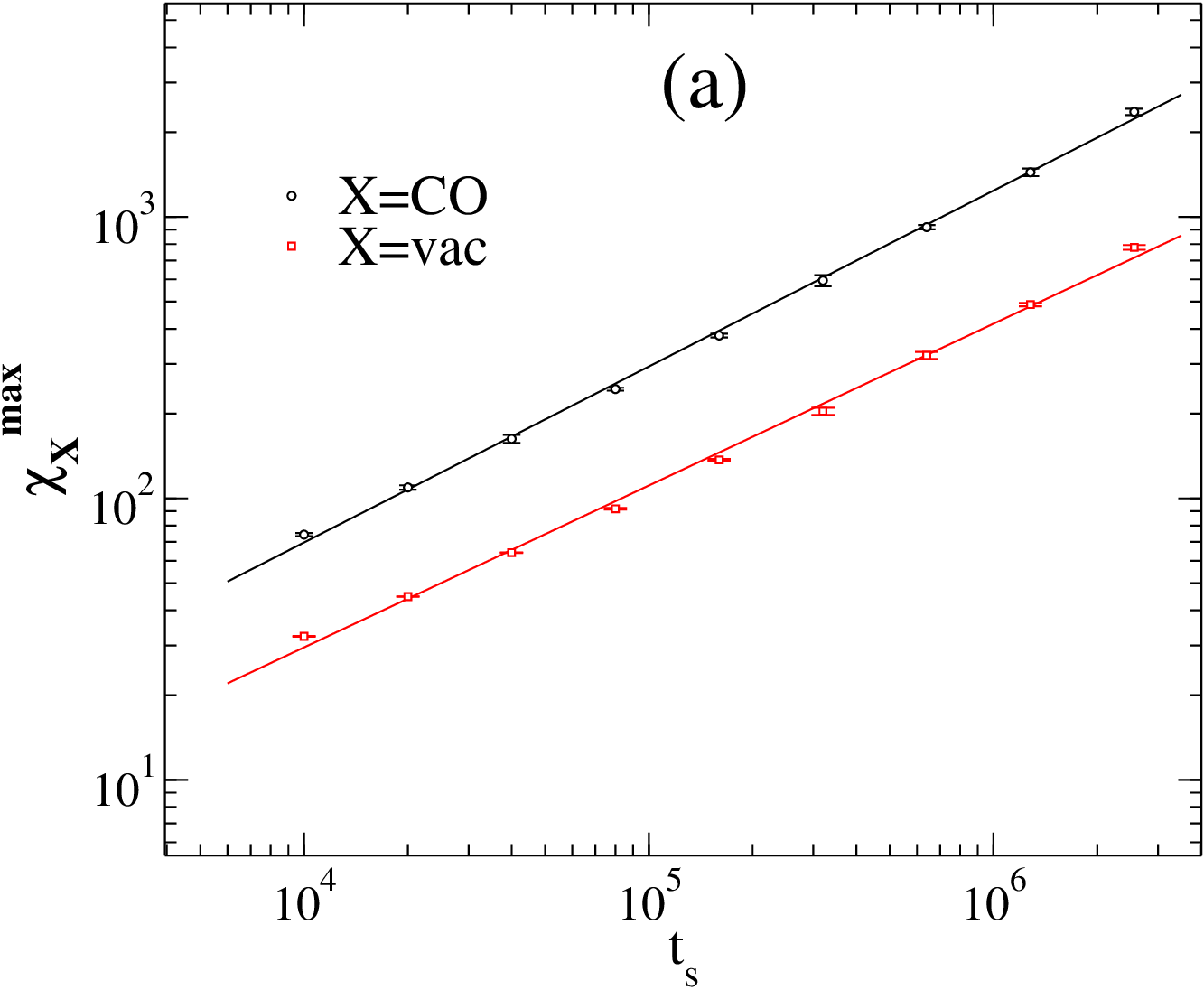}
       \\ 
       \includegraphics[width=5.0cm, bb=53 -20 550 550]{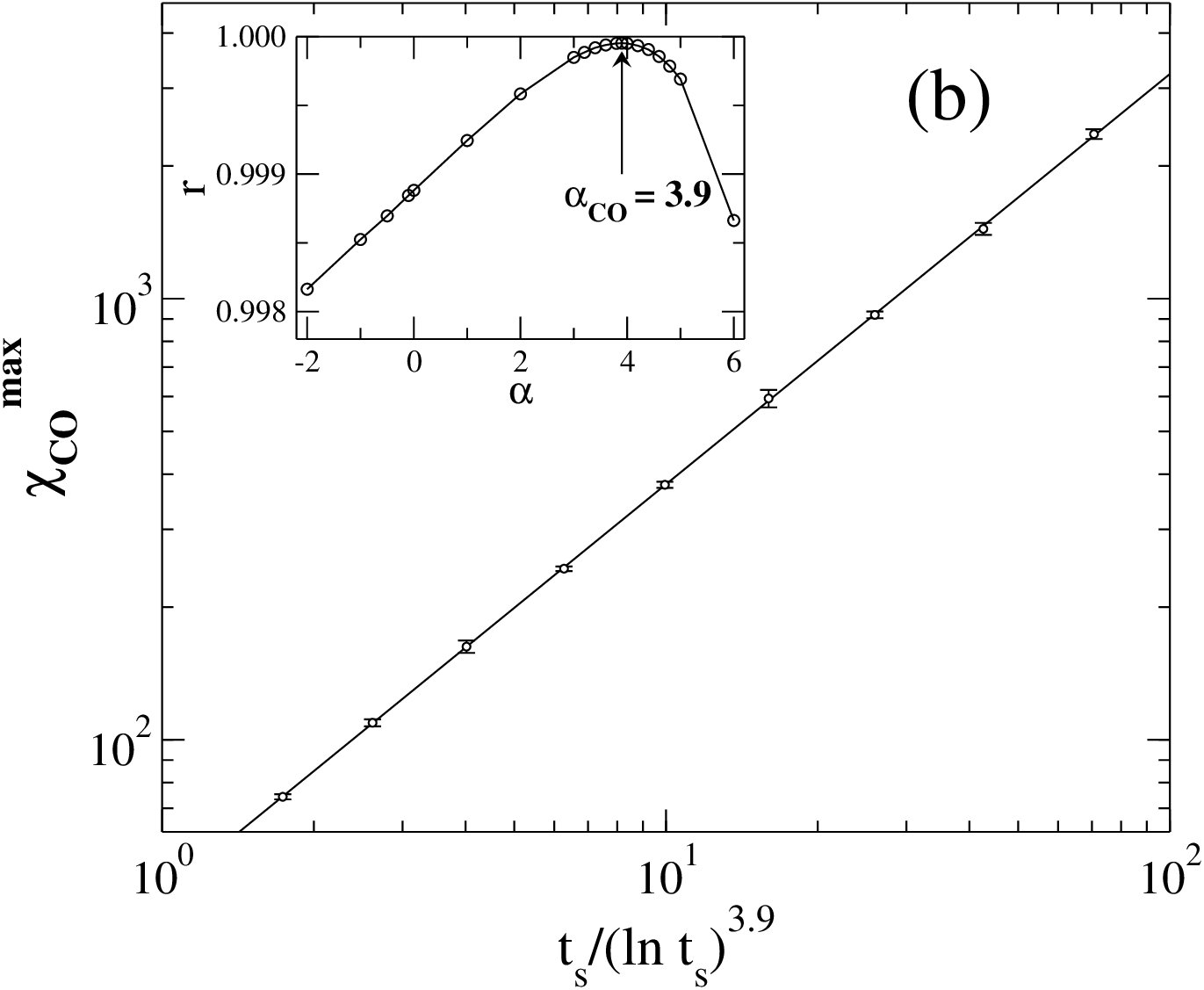}
       \\
       \includegraphics[width=5.0cm, bb=70 -20 550 550]{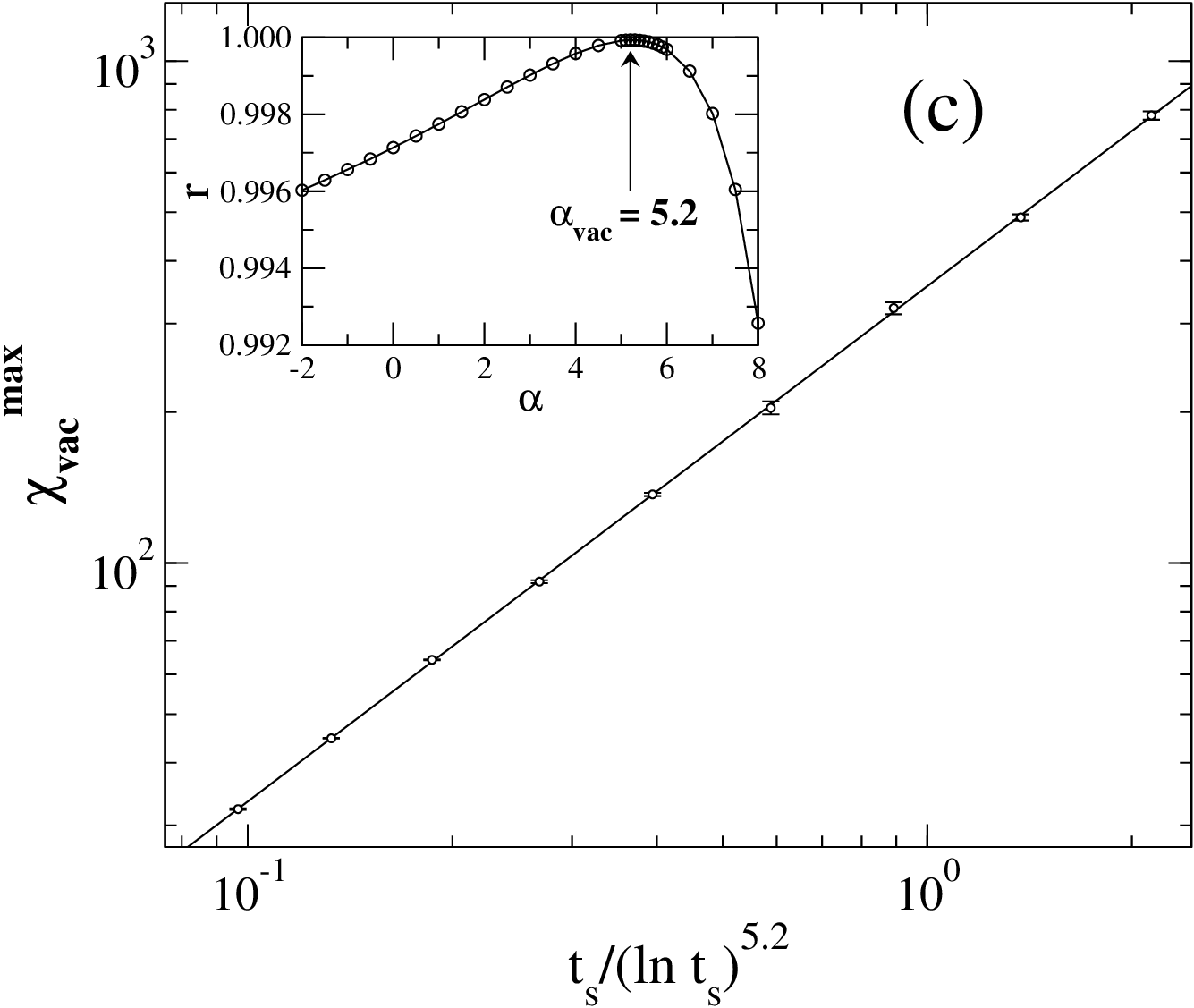}
     \end{tabular}
     \caption{(a) Maximum of the susceptibilities $\chi_{CO}$ and $\chi_{Vac}$ as a function of the time scale $t_s$ on a log-log scale. Straight lines are power law fits. 
     (b) Log-log plot of the maximum susceptibility $\chi^{max}_{CO}$ versus $t_s[ln(t_s)]^{-3.9}$. The inset shows the $r$-coefficient calculated for $\chi^{max}_{CO}$ versus $t_s [log (t_s)]^{-\alpha_{CO}}$ with maximum $r\approx 1$ at $\alpha_{CO}=3.9$. The straight line shows the fit of a power law with exponent $\beta_{CO}=0.93$.
     (c) Log-log plot of the maximum susceptibility $\chi^{max}_{vac}$ versus $t_s[ln(t_s)]^{-5.2}$. The inset shows the $r$-coefficient calculated for 
     $\chi^{max}_{vac}$ versus $t_s [log (t_s)]^{-\alpha_{vac}}$ with maximum $r\approx 1$ at $\alpha_{vac}=5.2$. The straight line is the fit of a power law with exponent $\beta_{vac}=1.03$. 
     }
     \label{fig.ChiMax}
   \end{center}
\end{figure}

Another signature of a phase transition is the increase in the maximum susceptibilities when the 
time scale $t_s$ increases, as shown in Figs. \ref{fig.1}(b) and \ref{fig.2}(b). 
Figure \ref{fig.ChiMax}(a) shows the maximum susceptibilities $\chi^{max}_{CO}$ and $\chi^{max}_{Vac}$ versus the time scale $t_s$ on log-log scales.
These maxima can be acceptably fitted with power laws given by 
\begin{equation}
\chi^{max}_{X} \propto t_s^{\beta_X}~~,
\label{Eq.purepowerlaw}
\end{equation}
where the subscript $X$ indicates the species $CO$ or $vac$. The straight lines in Fig. \ref{fig.ChiMax}(a) show power law fits of the maxima of each susceptibility, which give fitted exponents $\beta_{CO} =0.62 \pm 0.02$ and $\beta_{vac} =0.58 \pm 0.02$. 
In spite of the acceptable power law fit, a careful inspection of these figures reveals a systematic deviation of the data from the power law, suggesting that a pure power law could be improved with an additional factor. 
In the context of the finite-size scaling behavior in phase transitions, one of the most common additional factors is a logarithmic multiplicative factor of the linear size $L$ with its own exponent \cite{janke1997,salas1997,ruiz1998,janke2006,kenna2012}.
Analogously, here we propose that the maximum susceptibilities exhibit time scaling with a logarithmic multiplicative  factor:
\begin{equation}
\chi_{X}^{max} \propto t_s^{\beta_{X}} \left[log(t_s)\right]^{-\gamma_{X}}
\label{Eq.gamma}
\end{equation}
where $\gamma_X$ is a new exponent that accounts for the previous deviations.

In order to estimate the exponents $\beta_X$ and $\gamma_X$ of Eq. (\ref{Eq.gamma}), we will use the $r$-coefficient method as follows. We look for the best fit of a power law of $\chi^{max}_{X}$ versus $t_s [log (t_s)]^{-\alpha_X}$, varying the exponent $\alpha_X$. 
The insets of Figs. \ref{fig.ChiMax}(b) and \ref{fig.ChiMax}(c) show the $r$-coefficient calculated for these power laws when $\alpha_{CO}$ is varied from -2 to 6 and $\alpha_{vac}$ is varied from -2 to 8.
The maximum coefficient $r\approx1$ occurs for the exponents $\alpha_{CO}=3.9\pm0.3$ and $\alpha_{vac}=5.2\pm0.3$ indicated by vertical arrows in the insets of Figs. \ref{fig.ChiMax}(b) and  \ref{fig.ChiMax}(c), respectively. Therefore, these best fits for exponents $\alpha_X$ correspond to fitting the equation 
\begin{equation}
\chi^{max}_{X} \propto \{ t_s \left[log(t_s)\right]^{-\alpha_X} \}^{\beta_X}.
\label{Eq.beta}
\end{equation}
These fits are shown, on log-log scales, as straight lines in the plots of Figs. 
\ref{fig.ChiMax}(b) and \ref{fig.ChiMax}(c), with fitting exponents $\beta_{CO}=0.93\pm0.04$ and $\beta_{vac}=1.03\pm0.05$, respectively.
Finally, comparing Eq. (\ref{Eq.gamma}) with Eq. (\ref{Eq.beta}) gives us $\gamma_X=\beta_X \alpha_X$, and therefore we obtain the exponents $\gamma_{CO}=3.6\pm0.3$ and $\gamma_{vac}=5.3\pm0.4$.

In summary, the straight lines in Figs. \ref{fig.ChiMax}(b) and \ref{fig.ChiMax}(c) 
show the fits of Eq. (\ref{Eq.gamma}) for $\chi^{max}_{CO}$ with exponents $\beta_{CO}=0.93\pm0.04$, $\gamma_{CO}=3.6\pm0.3$, and  for $\chi^{max}_{vac}$
with exponents $\beta_{vac}=1.03\pm0.05$, $\gamma_{vac}=5.3\pm0.4$.
These fits are in excellent agreement with the simulation data, better than the fit of the pure power law given by Eq. (\ref{Eq.purepowerlaw}), shown as the straight lines in Fig. \ref{fig.ChiMax}(a). 

In this way, our proposed finite-time scaling given by Eqs. (\ref{Eq:pcocrit}),
(\ref{Eq.powerlawtheta}), and (\ref{Eq.gamma}) works very well in the range of the time scales studied in this work, supporting the assumption that $(p_c,\theta_c)$ behaves as a \textit{dynamic} phase transition point.

\begin{figure}
   \begin{center}
     \vspace{-0.5cm}
     \begin{tabular}{cc}
       \includegraphics[width=6.0cm, bb=70 -20 550 550]{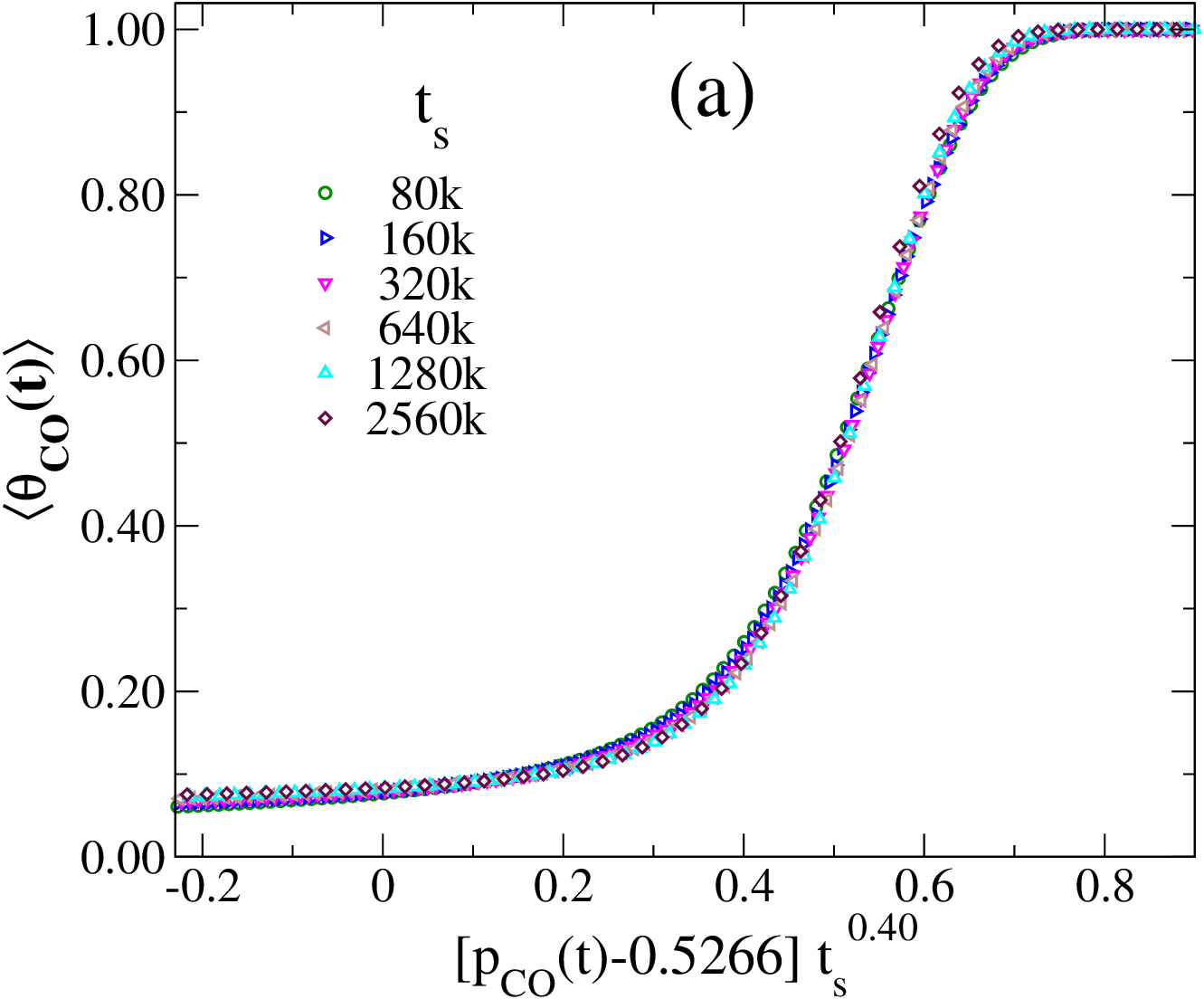}
       & \hspace{2.0cm}
       \includegraphics[width=6.0cm, bb=70 -20 550 550]{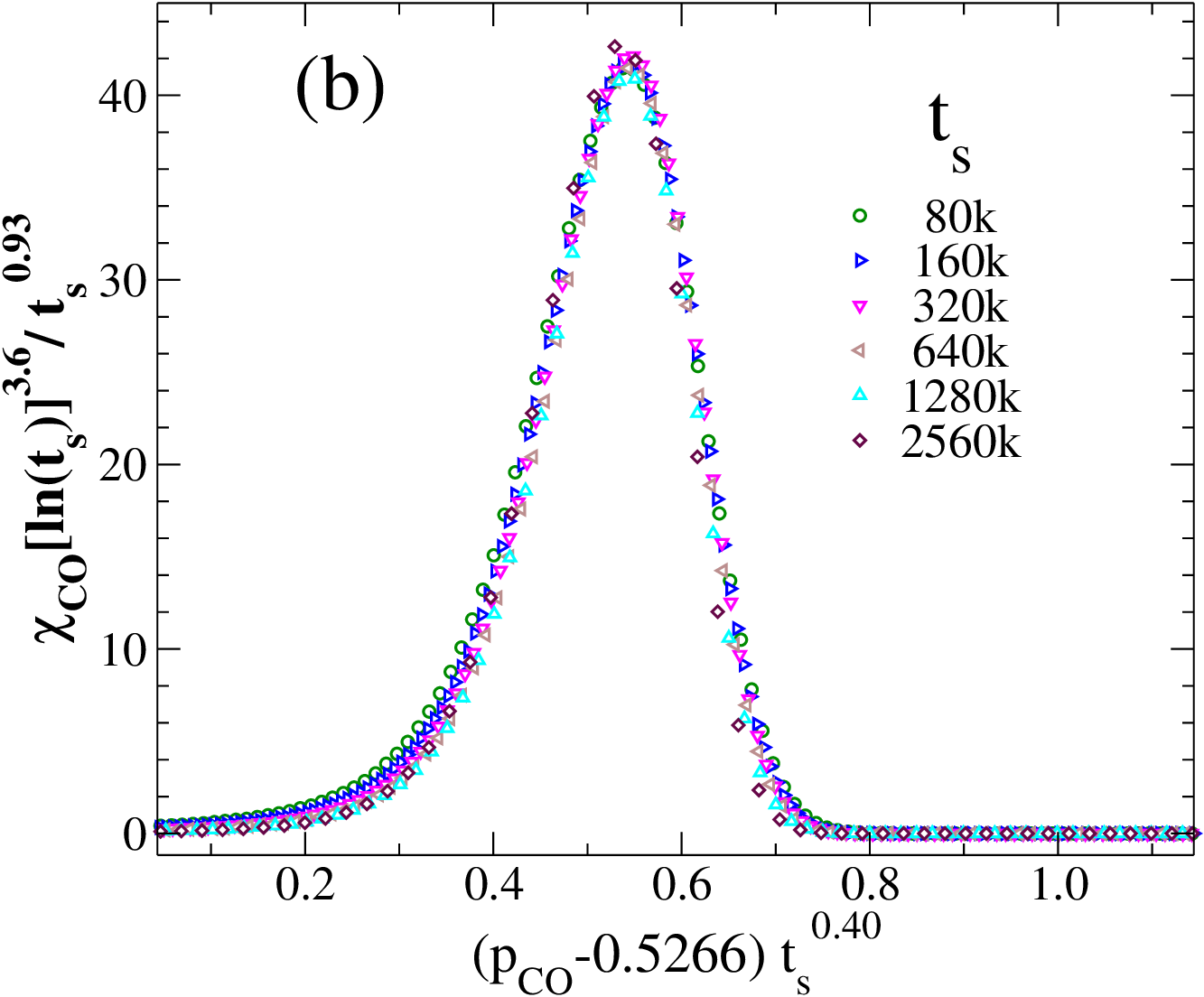}
       \\
       \includegraphics[width=6.0cm, bb=70 -20 550 550]{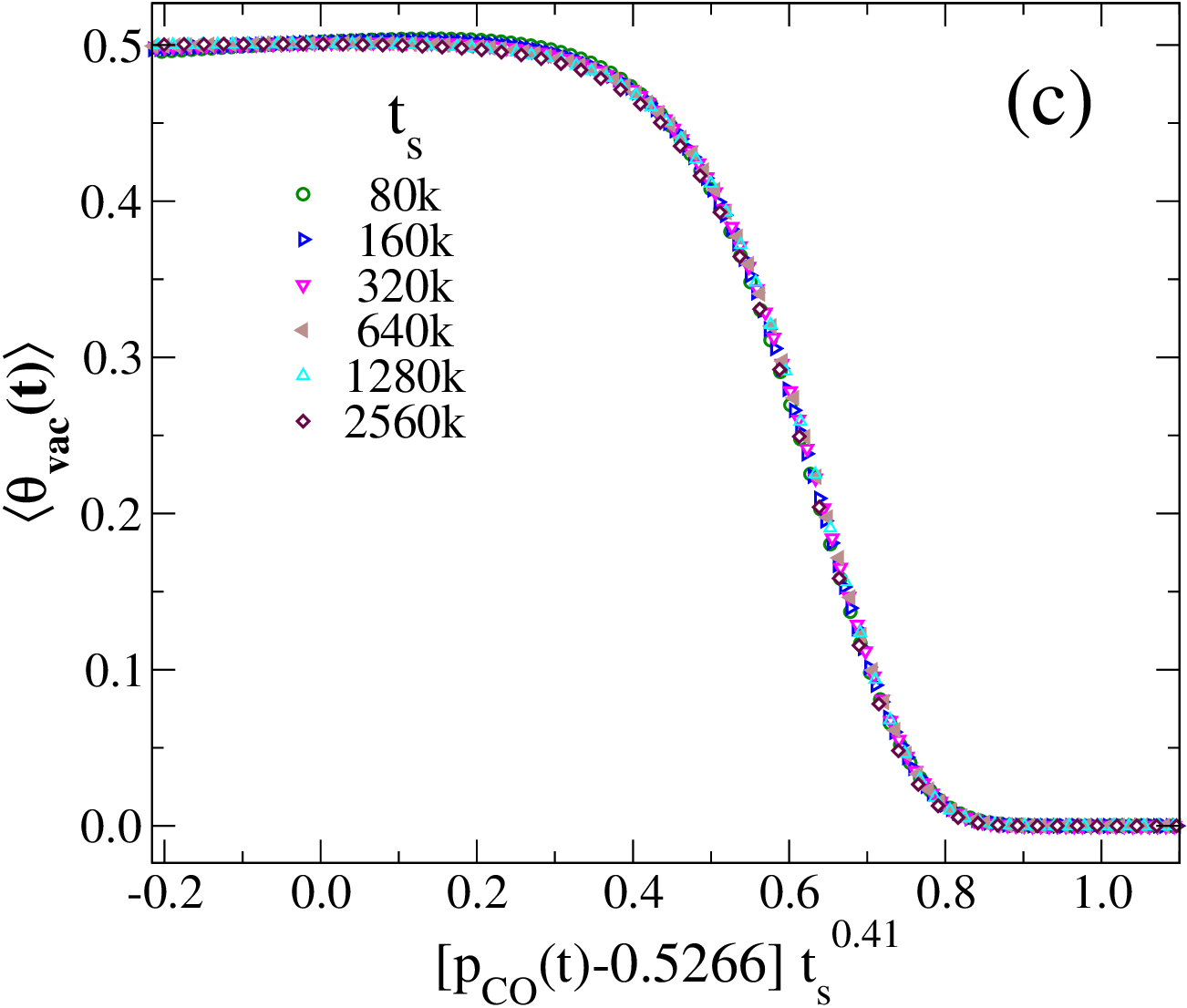}
       & \hspace{2.0cm}
       \includegraphics[width=6.0cm, bb=70 -20 550 550]{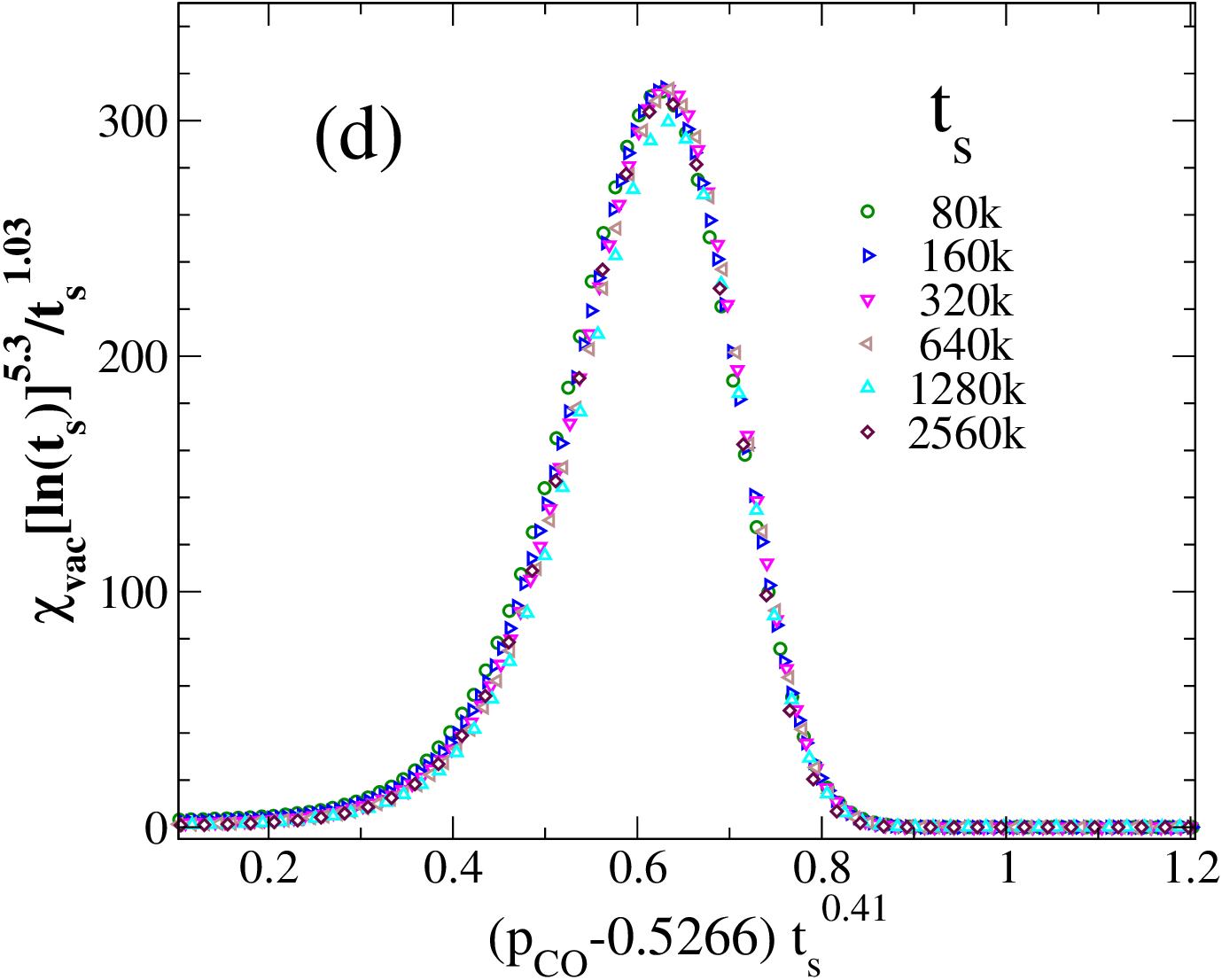}
     \end{tabular}
     \caption{(a) Collapses for the $CO$ coverage obtained rescaling the pressure $p_{co}$ by means of $(p_{CO}-0.5266)t_s^{0.40}$.
     (b) Collapses for the susceptibility $\chi_{CO}$ obtained rescaling the pressure $p_{co}$ by means of $(p_{CO}-0.5266)t_s^{0.40}$ and the maximun susceptibility by means of $\chi^{max}_{CO}[ln(t_s)]^{3.6}/t_s^{0.93}$.
     (c) Idem (a) with $(p_{CO}-0.5266)t_s^{0.41}$ for vacant sites. 
     (d) Idem (b) with $(p_{CO}-0.5266)t_s^{0.41}$ and $\chi^{max}_{vac}[ln(t_s)]^{5.3}/t_s^{1.03}$ for vacant sites.
     }
     \label{fig.collapse}
   \end{center}
\end{figure}
A more compelling test of the scaling laws given by Eqs. (\ref{Eq:pcocrit}) and
(\ref{Eq.gamma}) is achieved by collapsing the data of coverages and susceptibilities shown in Figs. \ref{fig.1} and \ref{fig.2}. 
In the case of coverages $\theta_{CO}$ and $\theta_{vac}$, we observe in Figs. \ref{fig.1}(a) and \ref{fig.2}(a) that they are similar to sigmoid curves as a function of $p_{CO}$.
Also, the position of the effective transition, given by the peak of the susceptibility, is governed by the power law (Eq. (\ref{Eq:pcocrit})), which reflects the typical size dependence using the time scale $t_s$ instead of the size $L$. Therefore, the collapse can be achieved using the pressure of the
transition $p_c$ and the exponent $\nu_X$, plotting the coverages versus the rescaled variable $(p-p_c)t^{1/\nu_X}$. 
Figure \ref{fig.collapse}(a) and \ref{fig.collapse}(c) show an acceptable data collapse, considering the data for the larger scale times $40$k$<t_s$, in the cases of the $CO$ coverage and vacant sites, respectively.

The scaling behavior of the dynamical susceptibilities can also be tested by plotting $\chi_{X}[\ln{t_s}]^{\gamma_X}/t_s^{\beta_X}$ versus $(p_{CO}-p_c)t_s^{1/\nu_X}$ and looking for data collapse. Figure \ref{fig.collapse}(b) shows an excellent collapse for $\chi_{co}$ using the transition pressure $p_c=0.5266$ with the exponents $1/\nu_{CO}=0.40$, $\beta_{CO}=0.93$, and $\gamma_{CO}=3.6$.
While in the case of the $\chi_{vac}$, Fig. \ref{fig.collapse}(d) shows the same rescaling using the transition pressure $p_c=0.5266$ and the exponents $1/\nu_{vac}=0.41$, $\beta_{vac}=1.03$, and $\gamma_{vac}=5.3$. Again, the data collapse is in good agreement with the scaling assumption
given by Eqs. (\ref{Eq:pcocrit}) and (\ref{Eq.gamma}).

All the data plotted in Fig. \ref{fig.collapse} show good collapses. This leads to 
independent controls and consistency checks of the transition value $p_c=0.5266$ 
and the exponents involved in the transition. 

\section{\label{sec:discussion}Discussion and Conclusions}
 
We have applied the DLP given by Eq. (\ref{Eq.protocol}) to the first-order IPT of the ZGB model. One consequence of applying this method is the onset of finite-time effects, which are analogous to the standard finite-size effects for phase transitions, as discussed in the previous section. 
Therefore, our results support the hypothesis of the existence of a transition point, which is a novel dynamic transition.
This dynamic transition should not be confused with the evaporation/condensation transition observed in both equilibrium models \cite{Binder1,Binder2,Martinos} and the ZGB model \cite{ccpnos,ccpnos2}. The latter appears in the ZGB model using the constant-coverage ensemble, in which the coverage of $CO$ is the control parameter and the pressure of $CO$ is the dependent variable
($p_{CO}(\theta_{CO})$). In this kind of simulation, the stationary (or history-independent) regime is sampled, and the obtained data correspond to the limit $t\rightarrow\infty$. Therefore, in these simulations, the order in which the limits are taken is schematically:   
\begin{equation}
\lim_{L\to\infty} \left[ \lim_{t\to\infty}  p_{CO}(\theta_{CO},L,t) \right ],
\label{Eq:limitsCC}
\end{equation}
in contrast to the order of the limits we used in this paper, given by Eq. (\ref{Eq:limits}). In the evaporation/condensation transition, there is a finite first-order transition with a gap in the pressure that disappears in the thermodynamic limit $L\rightarrow\infty$. Both the pressure and the coverage of the transition point tend to the usual first-order transition values $(\theta_2,p_2)$ \cite{ccpnos,ccpnos2}.

The pressure of the dynamic transition obtained in this work, with both coverage of $CO$ and empty sites, is $p_c=0.5266\pm0.0001$. These figures are very close
but significantly greater than the pressure corresponding to the first-order
IPT $p_2=0.525615\pm0.000005$ \cite{zgb,evans1991,ziff1992,ccpnos,gradiente,monetti2001}. 
We note that our determination of $p_c$, obtained by means of finite-time scaling, is intrinsically dynamic. Furthermore, the observed finite-time scaling should be a consequence of critical-like behavior. These two properties are the main signatures of {\it spinodal points}.
Due to the fact that a spinodal point theoretically corresponds to the limit of stability, its properties usually resemble the typical ones expected in systems exhibiting critical behavior.
For example, in equilibrium models, the spinodal corresponds to the extrapolation of the divergences of the susceptibility \cite{Herrmann1982}. These points can be associated with the percolation problem \cite{Klein1981} and with the fractal aspect of the largest cluster \cite{Trudu2006}. On the other hand, the technique called {\it short-time dynamics} \cite{STDreview}, which is in principle valid for critical points, has been shown to be useful for detecting these points in several different situations \cite{ZhengPotts,nos1,nos2,nos3,nos4}. 
These considerations lead us to believe that the dynamic transition point $(p_c,\theta_{c})$ we have determined in this work should be identified with the upper spinodal point corresponding to the first-order IPT of the ZGB model.
In fact, previous determinations of this upper spinodal point are in good agreement with our results.
In reference \cite{albanoSTD}, using the {\it short-time dynamics} technique, the value for the spinodal pressure $p_{sp}=0.52675\pm0.00005$ has been obtained with lattice sizes up to $L=512$. Also, in reference \cite{daSilva2016}, using the $r$-coefficient method, the value $p_{sp}=0.52738\pm0.00014$ has been found, which is slightly above our estimated pressure.
This difference may be due to size effects observed in \cite{daSilva2016}, 
since the reported value was obtained for a lattice of rather modest size ($L=320$), and 
the extrapolation to infinite size has not been done. In contrast, our estimated pressure, as discussed above, is determined in the thermodynamic limit.
Finally, in reference \cite{ccpnos2}, using the {\it dynamical} constant-coverage ensemble,
the spinodal point was identified as the point where the condition $dp_{CO}/d\theta_{CO}\equiv0$ 
is satisfied. Also, at this point, a critical distribution of droplets has been found.
The results obtained with this method and with lattice sizes up to $L=2048$ were 
the $p_{sp}=0.52670\pm0.00005$ and the spinodal $CO$-coverage $\theta_{sp}=0.091\pm0.001$. 

It is worth mentioning that the states generated with the protocol of slowly-varying 
uniform pressure near the dynamic critical point, where the dynamical susceptibility 
has a maximum, are quite different from the states obtained using 
other techniques, such as the constant-coverage ensemble \cite{ccpnos2} or the short-time dynamics technique \cite{albanoSTD,daSilva2016}.
The coverage of $CO$ obtained with the protocol of slowly-varying uniform pressure, at the dynamic transition point, is $\theta_{CO}\sim0.60$ (see, for example, Fig. \ref{fig.satur}). In contrast, in the region of interest, the $CO$-coverage obtained with simulations using the CC-ensemble or the short-time dynamics technique, is $\theta_{CO}\sim0.07$. Therefore, the present DLP generates new states that were not observed in previous simulations.

We have numerically explored the physical consequences of applying the DLP given by Eq. (\ref{Eq.protocol}) to the first-order irreversible phase transition of the ZGB model.
To the best of our knowledge, this is the first study applying the DLP to 
a far-from-equilibrium transition. 
The most important consequence is the onset of dynamic scaling expressed in terms
of the characteristic time $t_s$, valid in the thermodynamic limit and large $t_s$. 
The finite-time effects, just like the finite-size effects, allow us to determine 
very precisely, from numerical simulations, a dynamic critical point and associated exponents. 
This dynamic critical point given by $p_c=0.5266(1)$ and $\theta_c=0.087(1)$
coincides with the spinodal point previously determined in references
\cite{albanoSTD,ccpnos2,daSilva2016}.
In summary, our results support the hypothesis that the spinodal point can be detected using the DLP as a dynamic critical point.

\acknowledgments

This work was supported by Consejo Nacional de Investigaciones Científicas y Técnicas
(CONICET), Universidad Nacional de La Plata (Argentina),
and, Agencia Nacional de Promoción de la Investigación,
el Desarrollo Tecnológico y la Innovación (Agencia I + D + i).
Simulations were carried out on the computing cluster of the Unidad 
de Calculo, IFLYSIB (Argentina).

\end{document}